\documentclass[]{aastex631}
\usepackage{soul}

\newcommand{\kms}{\mbox{$\>{\rm km\, s^{-1}}$}}

\newcommand{\kpc}{\mbox{$\>{\rm kpc}$}} 
\newcommand{\pc}{\mbox{$\>{\rm pc}$}} 
\newcommand{\Gyr}{\mbox{$\>{\rm Gyr}$}}

\newcommand{\yr}{\mbox{$\>{\rm yr}$}}
\newcommand{\Msun}{\>{\rm M_{\odot}}}

\newcommand\degrees{^\circ}
\newcommand{\avg}[1]{\mbox{$\left<{#1}\right>$}}

\newcommand{\sig}[1]{\mbox{$\sigma_{#1}$}}

\newcommand{\feh}{\mbox{$\rm [Fe/H]$}}
\newcommand{\al}{\mbox{$\rm \alpha$}}
\newcommand{\ofe}{\mbox{$\rm [O/Fe]$}}
\newcommand{\alfe}{\mbox{$\rm [\alpha/Fe]$}}

\newcommand{\sfrd}{\mbox{$\Sigma_\mathrm{SFR}$}}
\newcommand{\rform}{\mbox{$R_\mathrm{form}$}}
\newcommand{\rfinal}{\mbox{$R_\mathrm{final}$}}
\newcommand{\tform}{\mbox{$t_\mathrm{form}$}}
\newcommand{\gaia}{{\it Gaia}}
\newcommand{\ges}{{\it Gaia}-Sausage-Enceladus}

\def\eg{{\it e.g.}}

\def\ie{{\it i.e.}}

\def\apjl{ApJ}% Astrophysical Journal, Letters
\def\apjs{ApJS}% Astrophysical Journal, Supplement
\def\aap{A\&A}% Astronomy and Astrophysics
\shorttitle{Twin-peaked bulge chemistry}
\shortauthors{Debattista et al.}
\graphicspath{{./}{figures/}}
\begin{document}

\title{The imprint of clump formation at high redshift. II.  The chemistry of the bulge}

\author[0000-0001-7902-0116]{Victor P. Debattista}
\affiliation{Jeremiah Horrocks Institute, University of Central Lancashire, Preston, PR1 2HE, UK}
\author{David J. Liddicott}
\affiliation{Jeremiah Horrocks Institute, University of Central Lancashire, Preston, PR1 2HE, UK}
\author[0000-0003-2478-6020]{Oscar A. Gonzalez}
\affiliation{UK Astronomy Technology Centre, Royal Observatory, Blackford Hill, Edinburgh, EH9 3HJ, UK}
\author[0000-0002-0740-1507]{Leandro {Beraldo e Silva}}
\affil{Department of Astronomy, University of Michigan, 1085 S. University Ave., Ann Arbor, MI 48109, USA}
\affil{Jeremiah Horrocks Institute, University of Central Lancashire, Preston, PR1 2HE, UK}
\author[0000-0002-7662-5475]{Jo\~ao A. S. Amarante}\altaffiliation{UCLan Visiting Fellow}
\affiliation{Institut de Ciencies del Cosmos (ICCUB), Universitat de Barcelona (IEEC-UB), Martí i Franquès 1, E-08028 Barcelona, Spain}
\affil{Jeremiah Horrocks Institute, University of Central Lancashire, Preston, PR1 2HE, UK}
\author{Ilin Lazar}
\affiliation{Centre for Astrophysics Research, School of Physics, Astronomy and Mathematics, University of Hertfordshire, Hatfield AL10 9AB, UK}
\author[0000-0002-5829-2267]{Manuela Zoccali}
\affiliation{Instituto de Astrof\'isica, Pontificia Universidad Cat\'olica de Chile, Av. Vicu\~na Mackenna 4860, 782-0436 Macul, Santiago, Chile}
\affil{Millennium Institute of Astrophysics, Av. Vicu\~na Mackenna 4860, 82-0436 Macul, Santiago, Chile}
\author[0000-0002-6092-7145]{Elena Valenti}
\affiliation{European Southern Observatory, Karl Schwarzschild\--Stra\ss e 2, D\--85748 Garching bei M\"{u}nchen, Germany}
\affil{Excellence Cluster ORIGINS, Boltzmann\--Stra\ss e 2, D\--85748 Garching bei M\"{u}nchen, Germany}
\author[0000-0003-0645-5260]{Deanne B. Fisher}
\affiliation{Centre for Astrophysics and Supercomputing, Swinburne University of Technology, P.O. Box 218, Hawthorn, VIC 3122, Australia}
\author[0000-0002-3343-6615]{Tigran Khachaturyants}
\affiliation{Jeremiah Horrocks Institute, University of Central Lancashire, Preston, PR1 2HE, UK}
\author[0000-0002-1793-3689]{David L. Nidever}
\affiliation{Department of Physics, Montana State University, P.O. Box 173840, Bozeman, MT 59717, USA}
\author[0000-0001-5510-2803]{Thomas R. Quinn}
\affiliation{Astronomy Department, University of Washington, Box
351580, Seattle, WA 98195, USA}
\author[0000-0001-9953-0359]{Min Du}
\affiliation{Department of Astronomy, Xiamen University, Xiamen, Fujian 361005, China}
\author[0000-0002-3838-8093]{Susan Kassin}
\affiliation{Space Telescope Science Institute, 3700 San Martin Drive, Baltimore, MD 21218, USA}
\affil{Johns Hopkins University, 3400 North Charles St., Baltimore, MD 21218, USA}
%
  


\begin{abstract}
In Paper I we showed that clumps in high-redshift galaxies, having a high star formation rate density (\sfrd), produce disks with two tracks in the \feh-\alfe\ chemical space, similar to that of the Milky Way's (MW's) thin$+$thick disks. Here we investigate the effect of clumps on the bulge's chemistry. The chemistry of the MW's bulge is comprised of a single track with two density peaks separated by a trough. We show that the bulge chemistry of an $N$-body$+$smoothed particle hydrodynamics clumpy simulation also has a single track. Star formation within the bulge is itself in the high-\sfrd\ clumpy mode, which ensures that the bulge's chemical track follows that of the thick disk at low \feh\ and then extends to high \feh, where it peaks. The peak at low metallicity instead is comprised of a mixture of in-situ stars and stars accreted via clumps. As a result, the trough between the peaks occurs at the end of the thick disk track.
We find that the high-metallicity peak dominates near the mid-plane and declines in relative importance with height, as in the MW. The bulge is already rapidly rotating by the end of the clump epoch, with higher rotation at low \alfe. Thus clumpy star formation is able to simultaneously explain the chemodynamic trends of the MW's bulge, thin$+$thick disks and the Splash.
\end{abstract}



%%%%%%%%%%%%%%%%%%%%%%%%%%%%%%%%%%%%%%%%%%%%%%%%%%%%%%%%%%%%%%%%%%%%%%%%%%%%%%

\section{Introduction}
\label{s:intro}

The chemistry of the Milky Way's (MW) bulge provides important clues about its formation. The early measurements of \citet{rich88}, \citet{mcwilliam_rich94} established that the bulge's metallicity distribution function (MDF) is broad, reaching supersolar metallicities.  More recent observations have shown that the MDF is at least bimodal with possible hints of additional peaks \citep{ness+13a, schultheis+17, rojas-arriagada+20, johnson+22}, although this may partly be due to fitting multiple Gaussians to an intrinsically skewed distribution.  Spectroscopic surveys such as ARGOS \citep{argos}, GIBS \citep{gibs} and APOGEE \citep{apogee} have mapped the chemistry across the bulge \citep[e.g.][]{ness+13a,  gonzalez+15b, zoccali+17, queiroz+21} generally finding that its \feh-\alfe\ plane exhibits a single track, with two peaks and a trough between them.  In contrast, in the Solar Neighborhood, two tracks\footnote{Different authors prefer either the term {\it tracks} or {\it sequences} to refer to the same thing. Throughout we will refer to {\it tracks}.} are evident: at fixed \feh, a high-\alfe\ track corresponds to the thick disk and a low-\alfe\ track corresponds to the thin disk. The bulge chemistry follows the thick disk track at low metallicity \citep{melendez+08, bensby+10, alves-brito+10, hill+11, bensby+13}, but then extends to the most metal-rich thin-disk stars.  The location of the knee in the \feh-\alfe\ plane has generally been found to be identical between the bulge and thick disk \citep{jonsson+17b, zasowski+19}, with perhaps minor differences \citep{johnson+14, bensby+17, schultheis+17}, which may be partly attributed to comparing bulge giants with local thick disk dwarfs. \citet{aawilliams+16} found bimodalities in the bulge's \feh\ and \alfe\ in the \gaia-ESO data, with the metal-rich stars exhibiting lower velocity dispersions than the metal-poor ones. The advent of the large APOGEE DR17 dataset, and matching data from \gaia\ Data Release 2 (DR2), have permitted more detailed studies of the bulge chemistry. \citet{lian+20} used the bulge's chemistry to model its star formation history (SFH) and concluded that it is comprised of three phases: an early high star formation rate (SFR) phase, which is interrupted by a quenched phase, which produces a gap in the chemistry, followed by a later secular phase of low SFR.

The chemistry of the disk(s) differs from these trends. Many explanations have been advanced for the disk \al-bimodality. The ``two-infall'' model of \citet{chiappini+97} \citep[see also][]{chiappini09, bekki_tsujimoto11, tsujimoto_bekki12, grisoni+17, khoperskov+21, spitoni+21} suggests that a high SFR episode formed the high-\al\ track, followed, around $8 \Gyr$ ago, by a drop in the SFR and then the infall of pristine gas that diluted the overall metallicity of the MW, giving rise to the low-\al\ population.
Recent work has focused on forming multiple chemical tracks via some variant of accretion events \citep{snaith+16, grand+17, mackereth+18, buck20}, including those of stars born out of the plane of the disk \citep{vintergatanI}.

In \citet[][hereafter Paper I]{clarke+19} we presented a simulation of an isolated galaxy that produced a disk chemical dichotomy similar to the MW's chemical thin$+$thick disks.  At early times (largely over the first $2\Gyr$, but continuing to $4\Gyr$ at a lower rate) the model develops clumps with high SFR densities, \sfrd.  The masses and SFRs of the clumps in this model are comparable to those observed in high-redshift galaxies \citep[e.g.][]{guo+15, dessauges-zavadsky+17, guo+18, cava+18, huertas-company+20}. The clumps represent a second mode of star formation, separate from the usual distributed star formation, with high \sfrd, leading to two tracks in the chemical, \feh--\alfe, plane.  The rate of clump formation declines rapidly as the gas fraction drops, thereby resembling the two-infall model. In agreement with \citet{bournaud+09}, Paper I showed that clumps produce a geometric thick disk. The chemical and geometric properties of the thick disk formed this way are consistent with those of the MW \citep{beraldoesilva+20}. Moreover, \citet{amarante+20} showed that the resulting low angular momentum tail of the old stars is consistent with the ``Splash'' population in the MW \citep{pdimatteo+19b, splash}.

Paper I showed that some of the clumps sink to the center of the galaxy, where they contribute to the formation of a bulge.
While definitively determining if clumps are long lived enough to build bulges is challenging due to observational systematics \citep[see, for instance, the discussion in][]{bournaud+14}, observations of the stellar populations \citep[e.g.][]{guo+18, lenkic+21} and gradients of clump mass \citep{huertas-company+20, ambachew+22} suggest that at least some fraction of clumps likely do survive long enough to fall into the bulge. 
The chemistry of bulges formed with a significant contribution from clumps has not been studied extensively in the literature, despite frequent suggestions that bulges, including the MW's, may be partly built from clumps \citep[e.g.][]{nataf17, queiroz+21}.  Interestingly, \citet{immeli+04} found a bimodal distribution of [Mg/Fe] within the bulge of their clumpy chemodynamical model.  \citet{inoue_saitoh12} found a metal-rich bulge formed from clumps but did not study the chemistry in greater detail.  Therefore in this paper we study the consequences of star formation in a clumpy mode on the chemistry of the bulge.

The paper is organized as follows. Section~\ref{s:simulations} presents the simulations used in this paper. The chemistry, star formation, kinematics, and spatial variation of the model bulges are presented in Section~\ref{s:chemistry}. We discuss our results, and give a brief summary of the main results, in Section~\ref{s:discussion}.

%%%%%%%%%%%%%%%%%%%%%%%%%%%%%%%%%%%%%%%%%%%%%%%%%%%%%%%%%%%%%%%%%%%%%%%%%%%%%

\section{The Simulations}
\label{s:simulations}

We use the clumpy simulation of Paper I, as well as a control simulation that fails to produce long-lived clumps; both these models are described in \citet{beraldoesilva+20}. The subgrid physics of the two models differs only in the strength of the feedback employed. Both models are evolved from the same initial conditions, comprised of a cospatial hot gas corona and dark matter halo with Navarro-Frenk-White \citep{nav_etal_97_nfw} profiles. The dark matter halo has virial mass of $10^{12} \Msun$ and a virial radius $r_{200} \simeq 200 \kpc$.  The gas corona, which constitutes $10\%$ of the mass within the virial radius, starts with spin $\lambda = 0.065$ \citep{bullock_angmom+01}, and as it cools, via metal line cooling \citep{sshen+10}, it settles into a disk. Stars form from dense gas (density $>1 \mathrm{cm}^{-3}$) when the temperature drops below 15,000~K and the flow is convergent. Gas particles are not allowed to cool below the resolution limit by setting a pressure floor $p_{floor} = 3 G\epsilon^2\rho^2$, where $G$ is Newton's gravitational constant, $\epsilon$ is the softening length, set at $50\pc$, and $\rho$ is the gas particle's density \citep{agertz+09b}. The feedback via supernovae Types Ia and II uses the blastwave prescription of \citet{stinson+06}.  In the clumpy model, we couple $10\%$ of the $10^{51}$ erg per supernova to the interstellar medium as thermal energy.  In contrast, in the high-feedback model, $80\%$ of the feedback energy is coupled to the gas.  As shown in previous studies \citep{hopkins+12, genel+12, buck+17, oklopcic+17}, high feedback coupling inhibits the clumps, and \citet{beraldoesilva+20} show that in that case the geometric properties of the disk(s) do not resemble those of the MW. Feedback via asymptotic giant branch stars is also included.  Gas chemical and thermal diffusion uses the method of \citet{sshen+10}.

We evolve the models in isolation using a smooth particle hydrodynamics$+N$-body tree-code based on {\sc gasoline} \citep{gasoline}.  The initial models are comprised of $10^6$ particles in both the dark matter and gas components; both models form $\sim 2 \times 10^6$ stars.  The clumpy model forms clumps during the first $2\Gyr$, continuing at a lower rate to $4\Gyr$, as shown in Paper I. The final disk galaxy has a rotational velocity of $242\kms$ at the Solar Neighborhood, making it comparable to the MW (see fig.~2 of Paper I).  The high-feedback model evolves without forming any significant long-lived clumps. Henceforth we refer to the two models as the clumpy and high-feedback models. 

Neither of these two models forms a bar. The formation of a bar quenches star formation within most of the body of the bar \citep[e.g.][]{khoperskov+18}. In order to compare with the MW, we assume that the MW's bar formed at $t=6~\Gyr$ (which would make it $\sim 8~\Gyr$ old now).

%%%%%%%%%%%%%%%%%%%%%%%%%%%%%%%%%%%%%%%%%%%%%%%%%%%%%%%%%%%%%%%%%%%%%%%%%%%%%

\section{Bulge stellar populations}
\label{s:chemistry}

\subsection{The chemistry of the bulge}
\label{ss:bulgechem}

\begin{figure*}
\centerline{
\includegraphics[angle=0.,width=0.5\hsize]{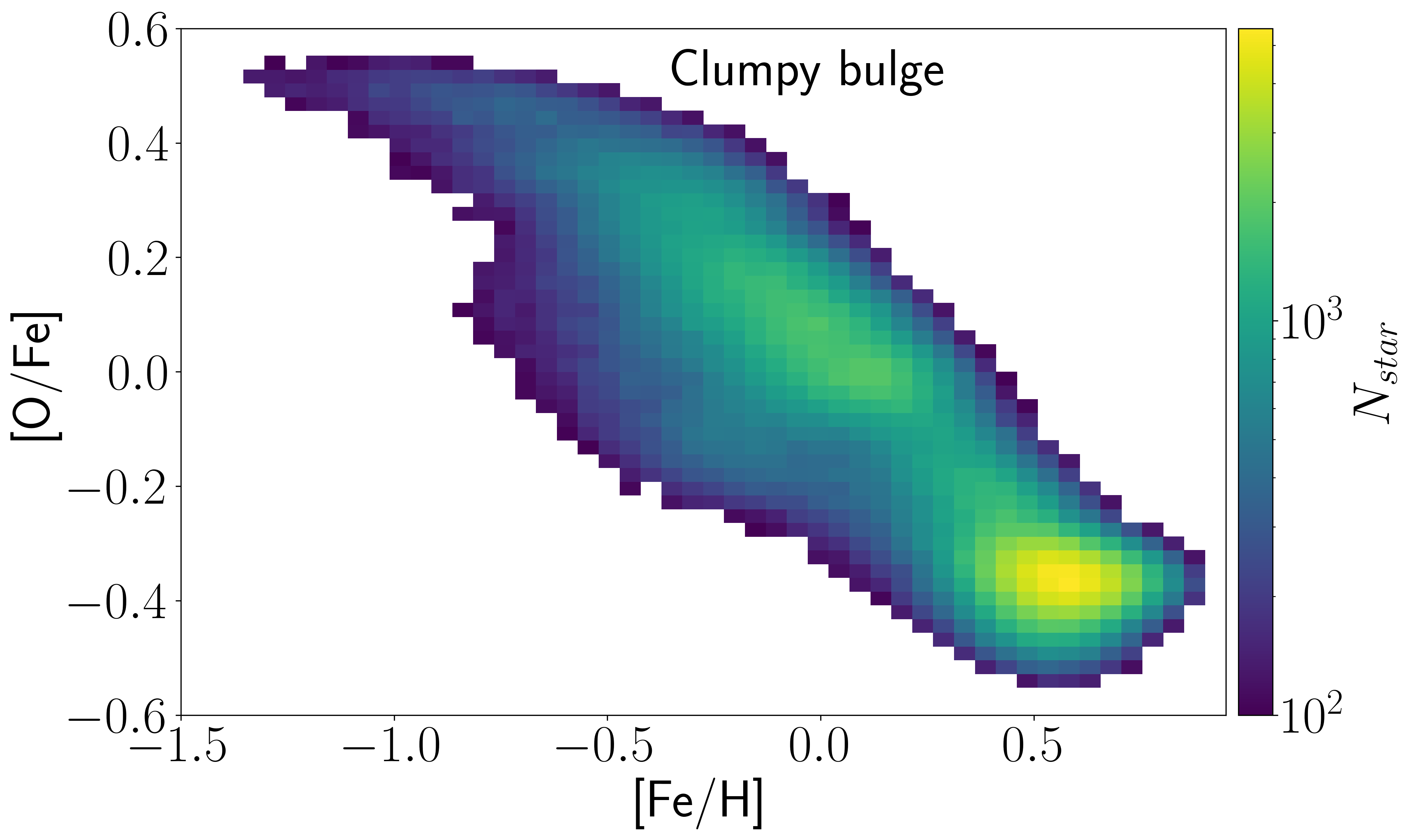}
\includegraphics[angle=0.,width=0.5\hsize]{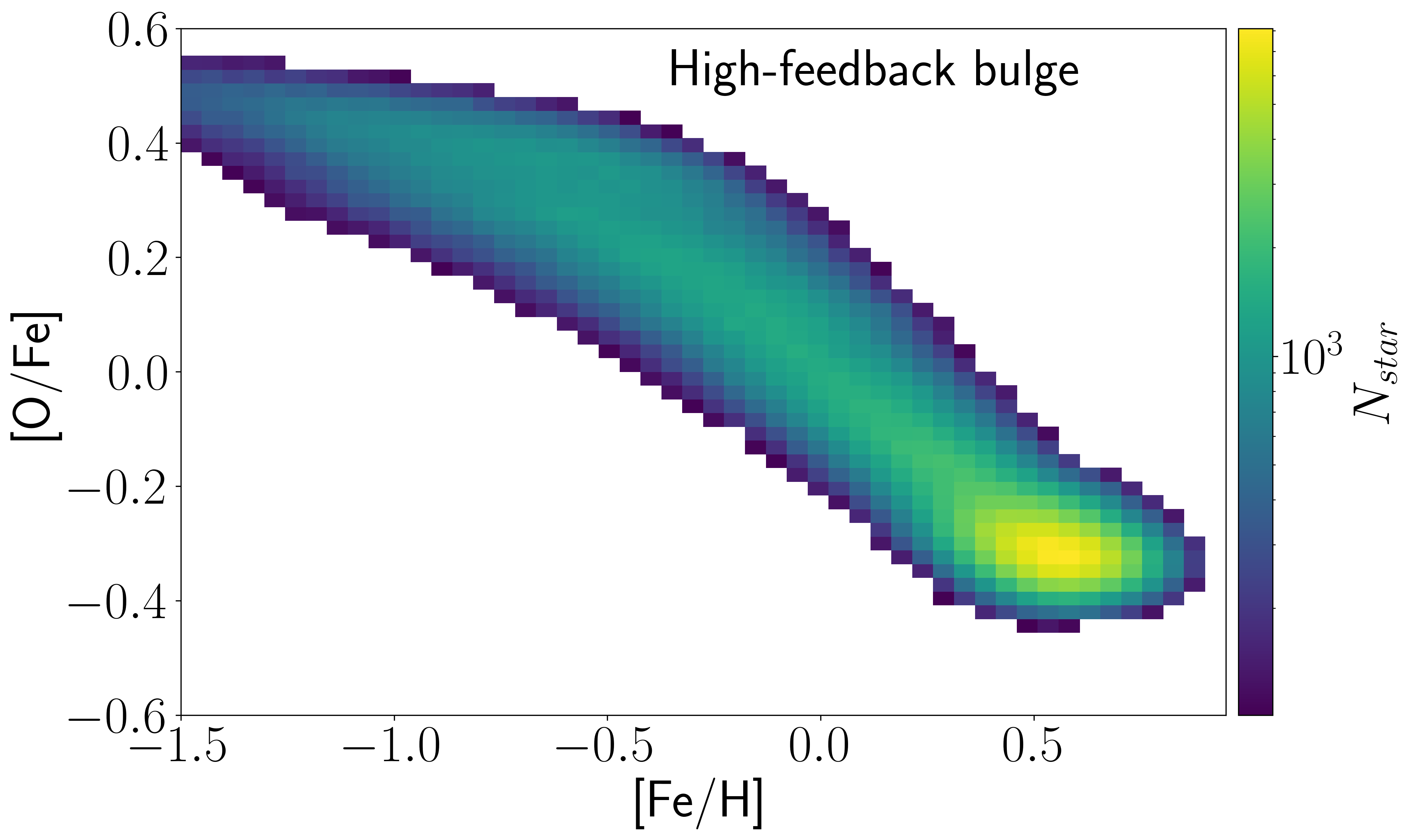}\
}
\centerline{
\includegraphics[angle=0.,width=0.5\hsize]{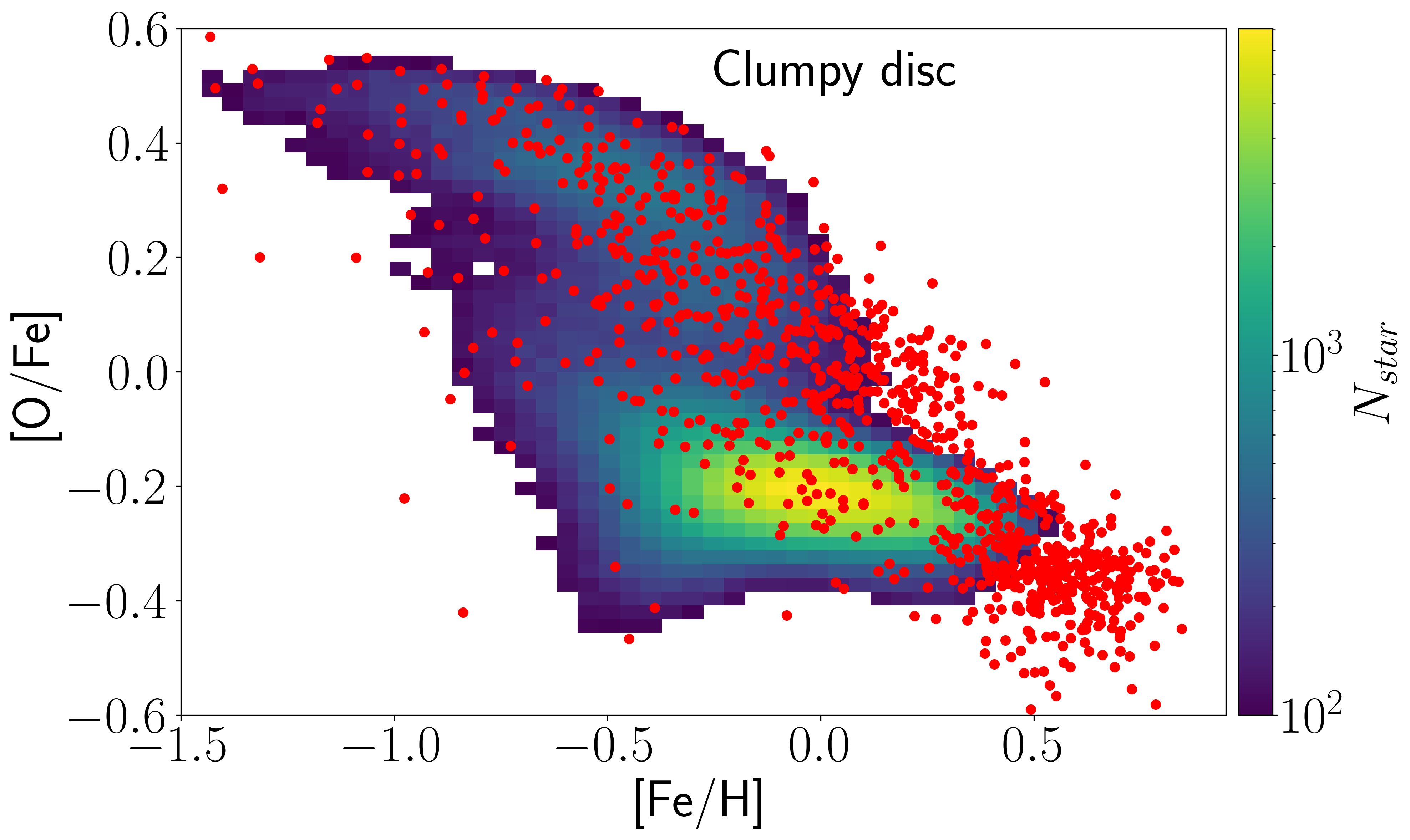}
\includegraphics[angle=0.,width=0.5\hsize]{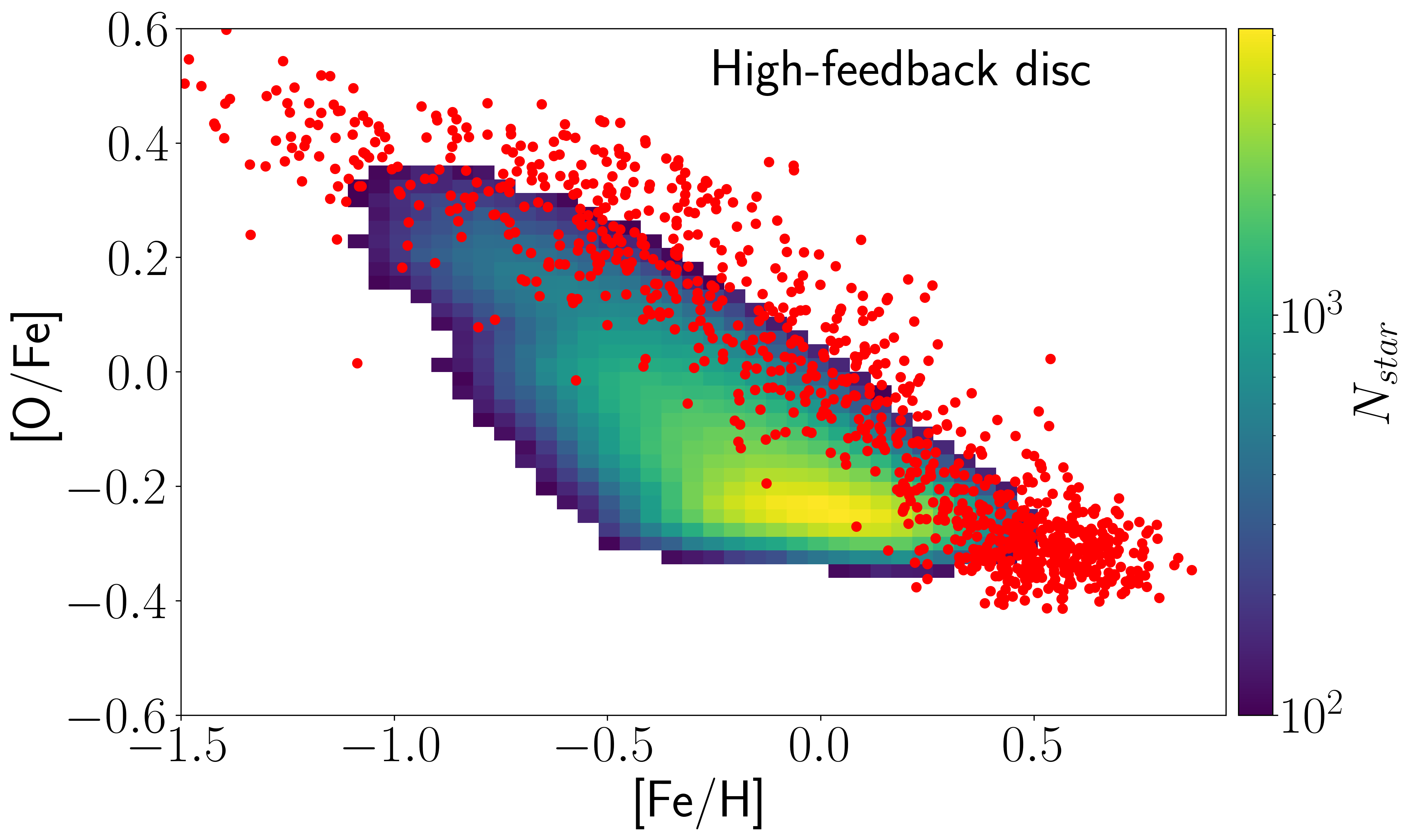}\
}
\caption{The density of stars in the \feh-\ofe\ chemical space at $t=10\Gyr$. Top: all stars within $R = 1~\kpc$.  Bottom: stars at $R > 5 \kpc$ with 1000 random bulge stars superposed as red points.  At left is the clumpy model, while at right is the high-feedback model. Smoothing in \feh\ and \ofe\ has been applied to all panels to match the chemical resolution of APOGEE, as described in Sec~\ref{ss:bulgechem}.  Bins with less than 100 stars have been suppressed. In the clumpy model, the bulge chemistry matches that of the disk in the high-\ofe\ region, in agreement with MW trends, and in contrast to the high-feedback model.
\label{f:chemicalspace}}
\end{figure*}

The top left panel of Fig.~\ref{f:chemicalspace} presents the chemistry of the stars within a galactocentric radius $R = 1~\kpc$ at $10~\Gyr$ in the clumpy model.  As in Paper I, we apply Gaussian measurement uncertainties of $\sig{\feh} = 0.1$ and $\sig{\ofe} = 0.03$ to mimic the measurement errors in APOGEE \citep{nidever+14}. The chemical space has a single track, with the density peaked at two locations: one metal-rich at $\feh \simeq 0.55$ and a broader metal-poor peak at $\feh \simeq -0.1$. The bottom left panel of Fig.~\ref{f:chemicalspace} presents the chemistry of the clumpy model's thin$+$thick disks at $R>5\kpc$, and compares this with the chemistry of the model's bulge (the red points represent a random selection of 1000 bulge particles). The chemistry of the bulge follows that of the thick disk at $\feh \la 0$, and then continues to more metal-rich than the thin disk.  The MW's bulge exhibits the same trend \citep[e.g.][]{melendez+08, bensby+10, alves-brito+10, hill+11, bensby+13, lian+20}. We have verified that the trends in Fig.~\ref{f:chemicalspace} are already in place by $t=6\Gyr$.

The right panels of Fig.~\ref{f:chemicalspace} present the chemistry of the high-feedback model. A number of important differences between the clumpy and high-feedback models are evident. The first difference is that the track of the bulge in chemical space no longer has two peaks. Instead the bulge has a single sharp peak at $\feh \simeq 0.6$ with a long tail to lower metallicities. Moreover, this model does not have a bimodal chemical distribution in the disk \citep[see also][]{beraldoesilva+20}, which happens because the high-\al\ stars form only via the clumpy star formation mode in these simulations. As a consequence, the bulge chemical distribution is offset vertically in \ofe\ relative to the disk. While the bulge has a high SFR and can therefore reach a high \ofe, this is not the case in the disk, and the bulge ends up more \al-rich than the disk. The lack of a trough in the bulge's chemistry and the difference between the bulge's peak \al\ and that of the disk are different from the trends observed in the MW.

\begin{figure}
\centerline{
\includegraphics[angle=0.,width=\hsize]{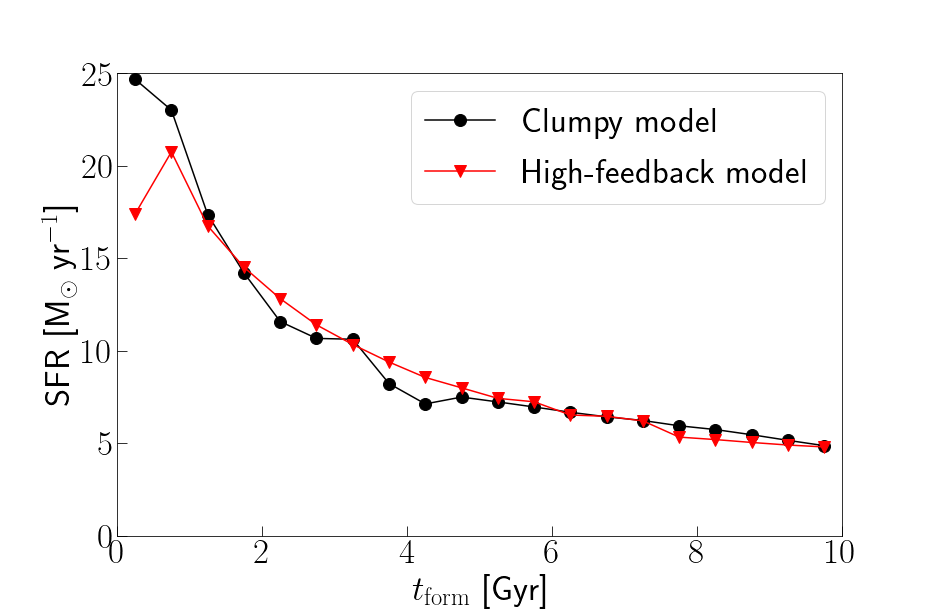}
}
\caption{The overall star formation history of the clumpy and high-feedback models.
\label{f:globalsfh}}
\end{figure}

In spite of these differences in chemical space, the overall SFH of these two models is very similar, as seen in Fig.~\ref{f:globalsfh}. The main difference is at early times, when the presence of the clumps briefly raises the overall peak SFR by $\sim 20\%$. In the high-feedback model, these clumps are short-lived \citep{genel+12, hopkins+12, buck+17, oklopcic+17}, and the SFR is therefore briefly lower.

\subsection{Evolution of the bulge's chemistry}
\label{ss:chemevol}

The fact that the clumpy model's bulge chemistry has a single, double-peaked track that matches that of the thick disk at $\feh \la 0$ is strikingly similar to what is observed in the MW. Understanding this trend therefore can help unravel the formation of the MW's bulge. Thus we next explore the evolution of the bulge chemistry to understand how clumpy star formation produces these trends.

\begin{figure*}
\centerline{
\includegraphics[angle=0.,width=0.5\hsize]{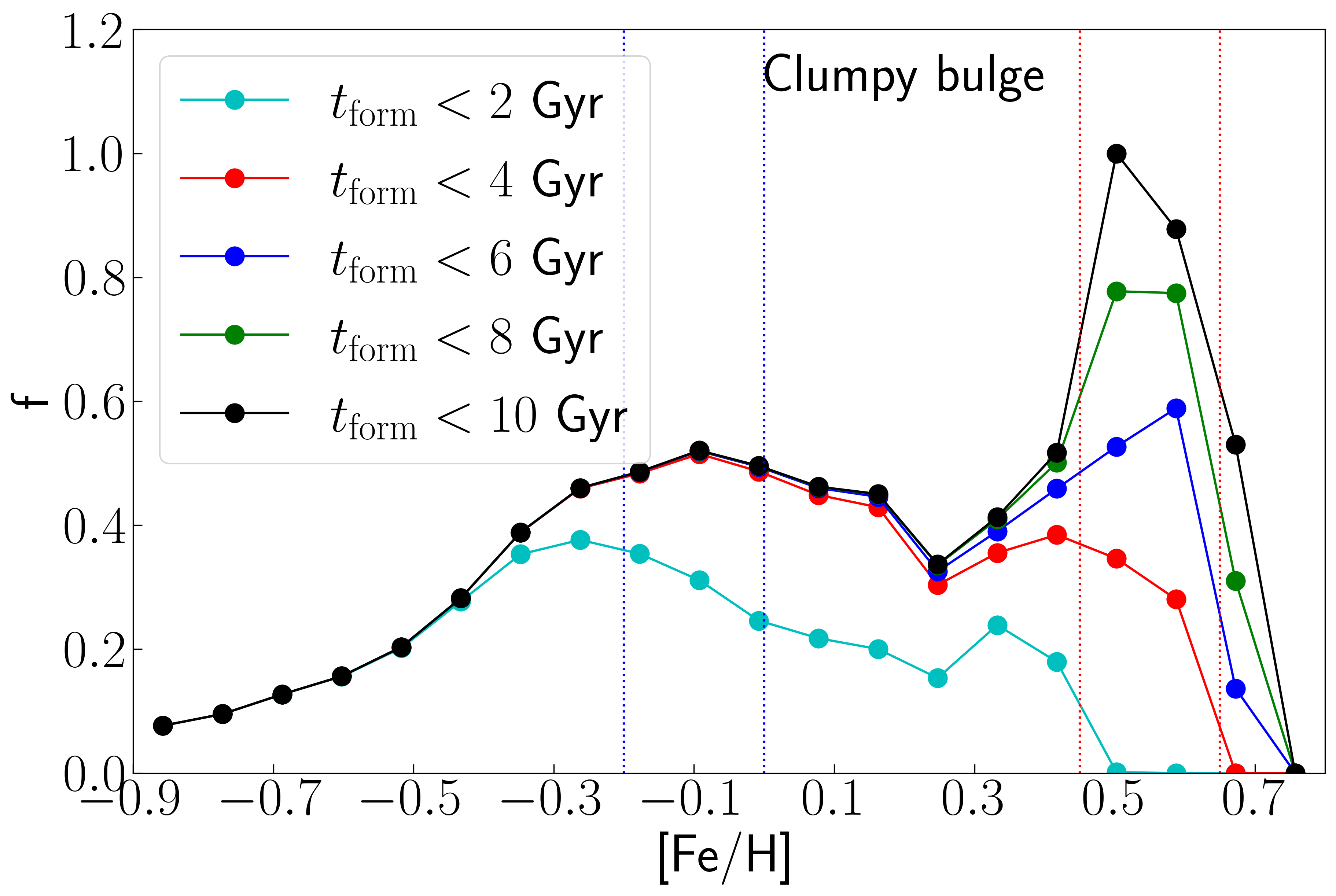}
\includegraphics[angle=0.,width=0.5\hsize]{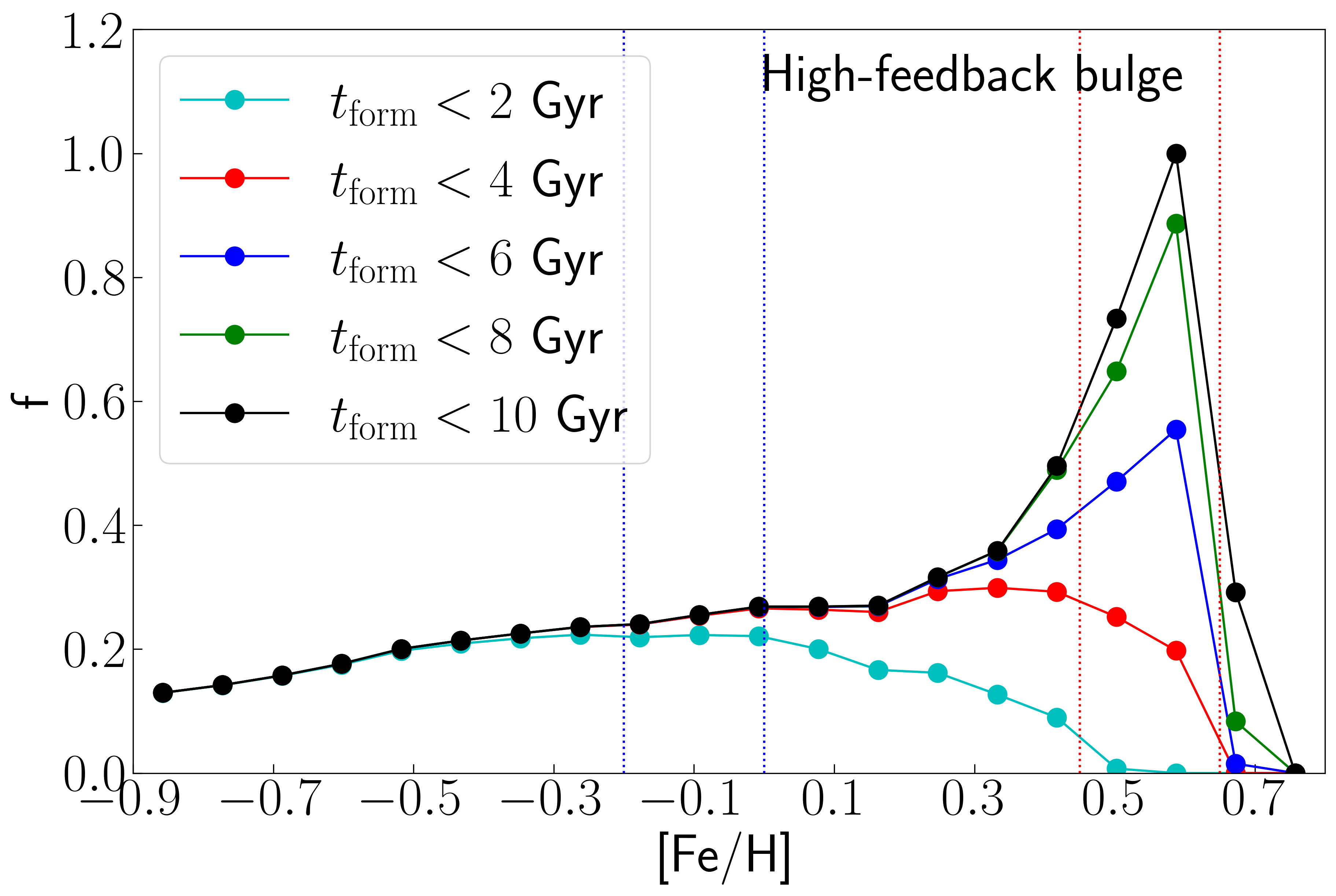}\
}
\centerline{
\includegraphics[angle=0.,width=0.5\hsize]{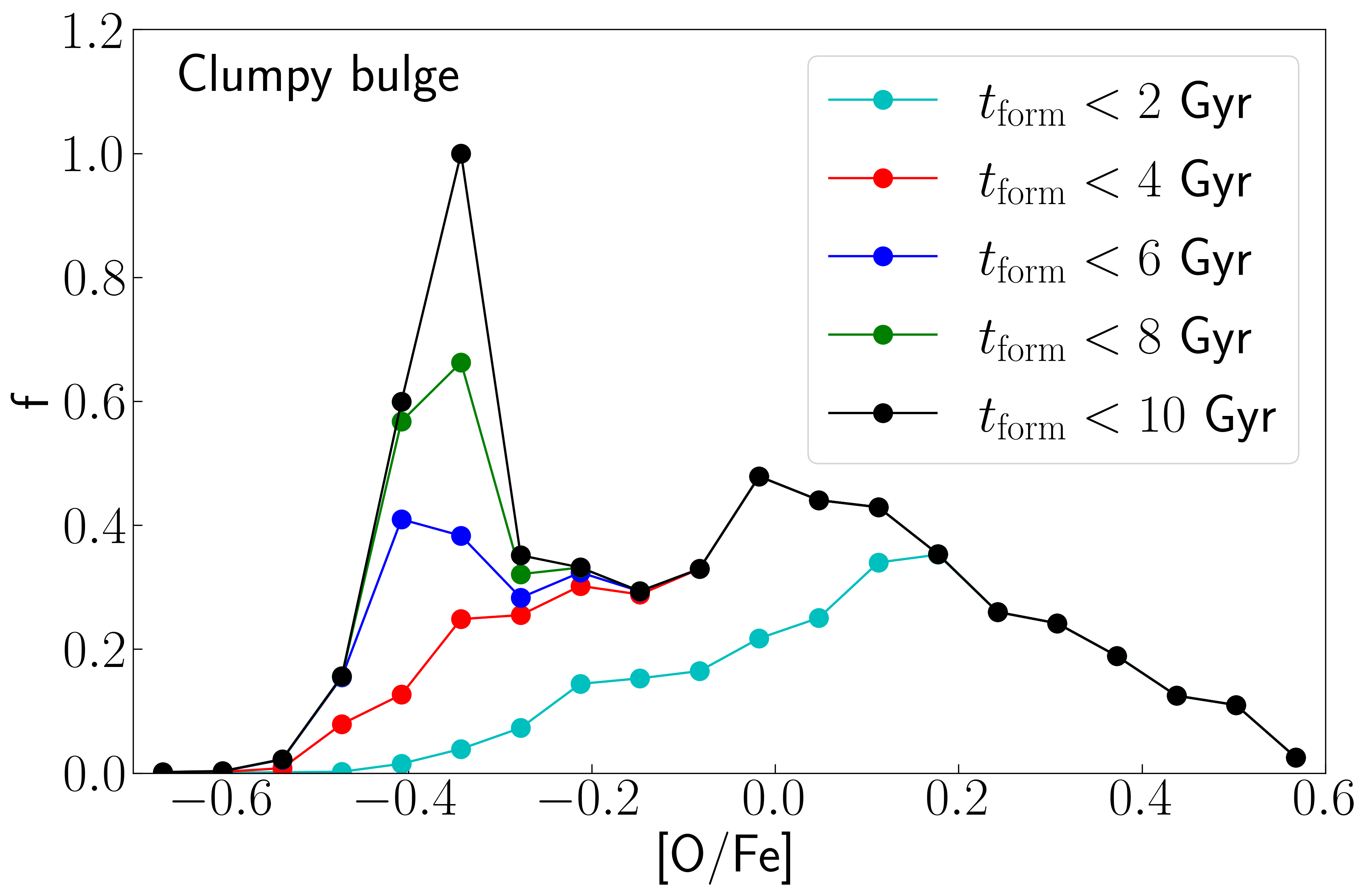}
\includegraphics[angle=0.,width=0.5\hsize]{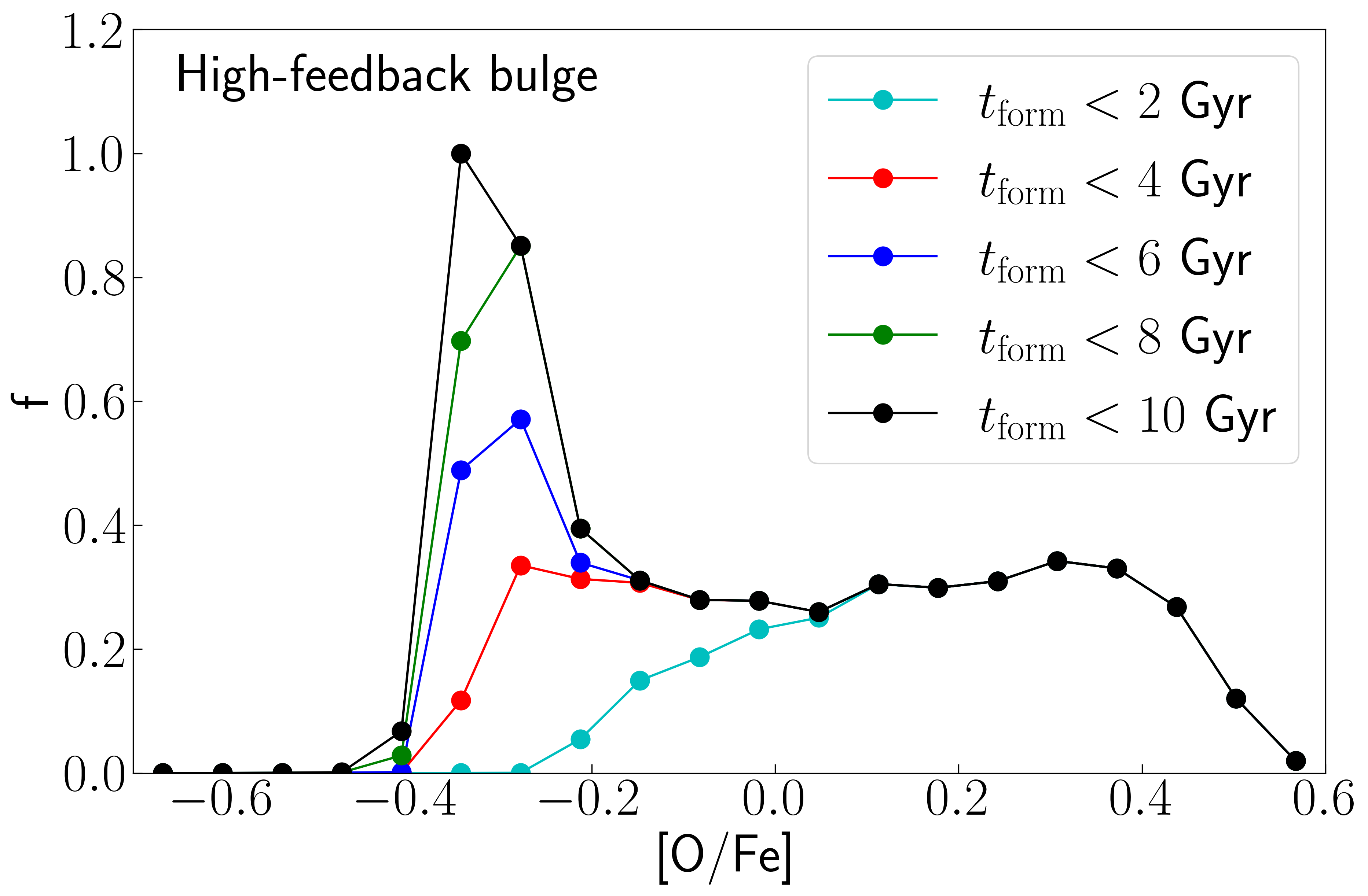}\
}
\caption{The evolution of the MDF (top) and \al DF (bottom) of stars  within the bulge ($R \leq 1\kpc$) between $t=2\Gyr$ and $t=10\Gyr$. At left is the clumpy model, and at right is the high-feedback one. In the clumpy model, a bimodality is present in the MDF at $t=10\Gyr$ with a broad, low peak at $\feh \sim -0.1$ and a narrow, high peak at $\feh \sim 0.5$. The bimodality is already evident, although weaker, at $t=2\Gyr$, when clump formation has started to die down, and is well established at $4\Gyr$. A bimodality is also present in the \al DF at $t=10\Gyr$, with a broad, low peak at $\ofe \sim 0$ and a narrow, high peak at $\ofe \sim -0.3$. This bimodality is significantly weaker and/or absent at $t=4\Gyr$. In the high-feedback model, instead, only a single peak develops in the MDF although the \al DF still has a weak second peak. All distributions have been normalized to the corresponding peak at $10\Gyr$. In the top row, the vertical dotted lines indicate the regions around the peaks where we define MDF peaks discussed in Section~\ref{ss:ages}.
\label{f:modf}}
\end{figure*}

Fig.~\ref{f:modf} shows the chemical evolution of the bulge inside $R = 1\kpc$ for both models. We show the MDF and the \al\ distribution function (\al DF) for all bulge stars formed up to 2, 4, 6, 8 and 10~\Gyr. The clumpy model, at $t=4\Gyr$, when clump formation fully ceases, has a bulge MDF which is bimodal (top left panel), with a low-metallicity peak at $\feh \simeq -0.1$ and a small peak at $\feh \simeq 0.4$.  The high-metallicity peak grows in importance as subsequent in-situ star formation adds a population of high-metallicity stars. The trough between the two peaks falls at $\feh \simeq 0.25$. In the MW's bulge, the metallicity of the trough varies with position in the range $\feh \sim -0.2$ to $0.2$ \citep{zoccali+17}.
After $t=4~\Gyr$, the \al DF of the clumpy model (bottom left panel) has a fixed peak at high \ofe\ (at $\approx 0$, but we caution that \ofe\ values often have significant offsets in simulations compared to observations, as we also found in Paper I.)  At $t=4\Gyr$, the \al DF has a point of inflection at low \ofe, where a pronounced second peak later develops. A double-peaked \al DF is similarly present in the MW's bulge \citep[e.g.][]{lian+20}

In contrast, the chemical evolution of the high-feedback model (right panels) results in only a single peak in the bulge's MDF, and only a weak double peak in the bulge \al DF. At best a weak trough is visible in chemistry of its bulge. The two models differ at the low-\feh\ peak (\ie\ at the high-\ofe\ peak), which must represent the location where the clump formation plays an important role in one model and is absent from the other.

Small differences between the clumpy and the high-feedback models
  are already present at $2\Gyr$, which Fig.~\ref{f:globalsfh} shows
  has the largest differences between the global SFRs of the two
  models. At $2\Gyr$ the MDF of the clumpy bulge has a peak at low
  \feh, while a peak at high \feh\ is incipient, but not yet
  prominent. The high-feedback bulge has a very similar MDF, but it has
  only a single peak at roughly the same subsolar \feh\ as in the
  clumpy model. The low-\feh\ peak is more prominent in the clumpy
  bulge than that in the high-feedback bulge, but the overall trends are
  similar.  Similarly the \al DFs of the two models are not yet very
  different, with a single peak at high \al.  The differences between
  the chemistry of the two bulges become larger between $2$ and
  $4\Gyr$, despite the fact that the global SFRs of the two models are more
  similar at these times. In the clumpy model, the separate peak at
  high \feh\ now becomes more developed, while the continuing
  enrichment in the bulge of the high-feedback model results in only a
  single peak at high \feh. The low-\feh\ peak in the clumpy model
  grows in importance at this time, while shifting to higher \feh. At
  the same metallicities as the low-\feh\ peak of the clumpy model,
  the bulge of the high-feedback model barely changes during this
  time. In the high-feedback bulge, the \al DF begins to develop a peak
  at low \alfe, while in the clumpy bulge the low-\alfe\ peak has not
  yet started to be visible, but the high-\al\ peak continues to grow
  while shifting to lower \alfe.  As we show below, the driver of
  these differences is the infall of clumps into the bulge of the
  clumpy model between $2$ and $4\Gyr$.  After $4\Gyr$, when no
  further clumps form in the disk of the clumpy model, the chemical
  evolution of the two bulges proceeds very similarly, with an increasing
  numbers of stars at the high-\feh, low-\alfe\ peaks. During this
  time, the clumpy model develops a second peak at low \alfe, which had
  formed earlier in the high-feedback model. In summary, it is not the
  differences in their SFRs that give rise to the different
  chemistries of the two bulges, but the infall of clumps onto the
  bulge of the clumpy model, which drives the continued growth of the
  low-\feh\ peak in its chemistry.

\subsection{Formation location}
\label{ss:formlocation}

\begin{figure*}
\centerline{
\includegraphics[angle=0.,width=0.48\hsize]{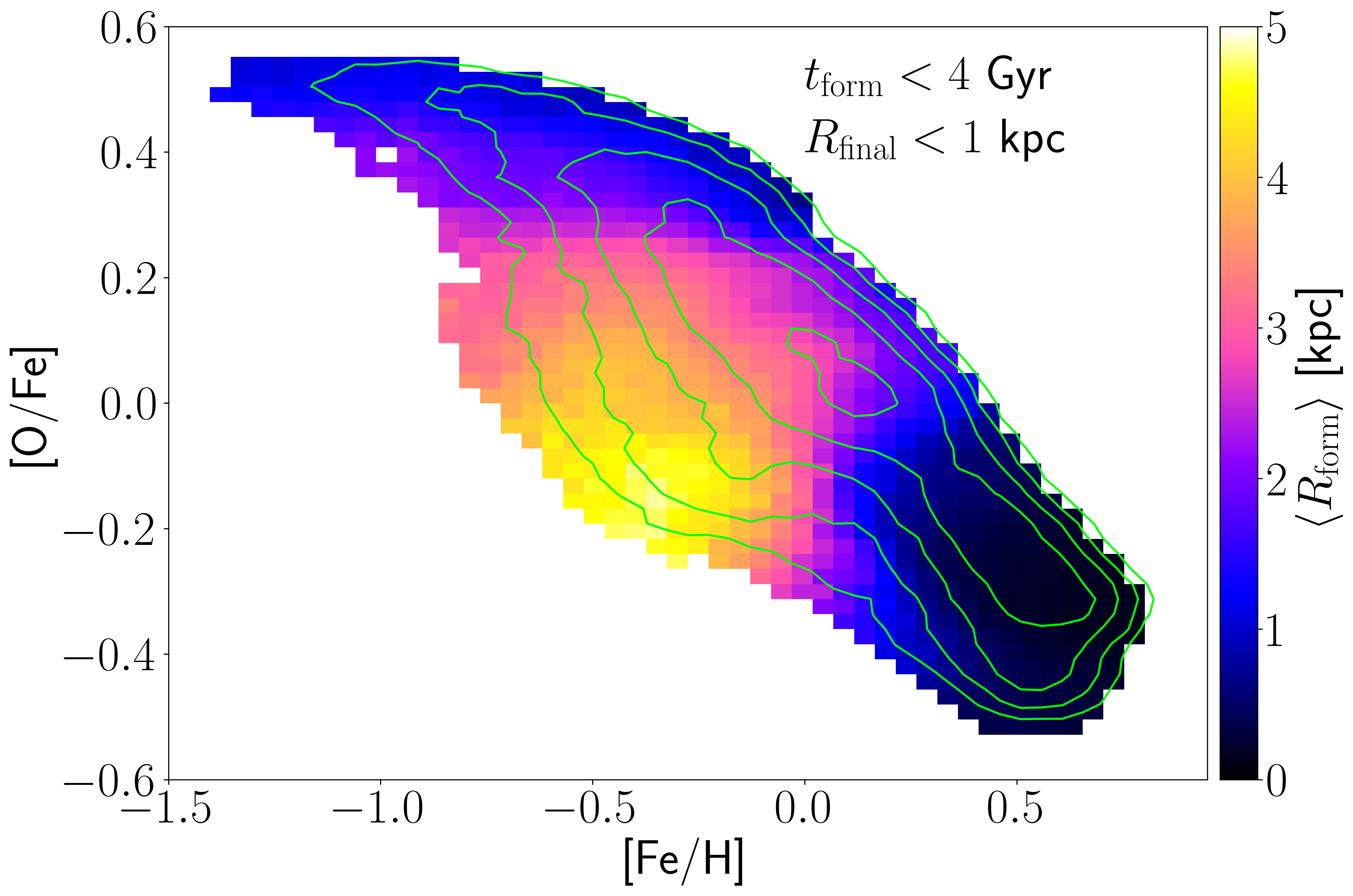}
\includegraphics[angle=0.,width=0.5\hsize]{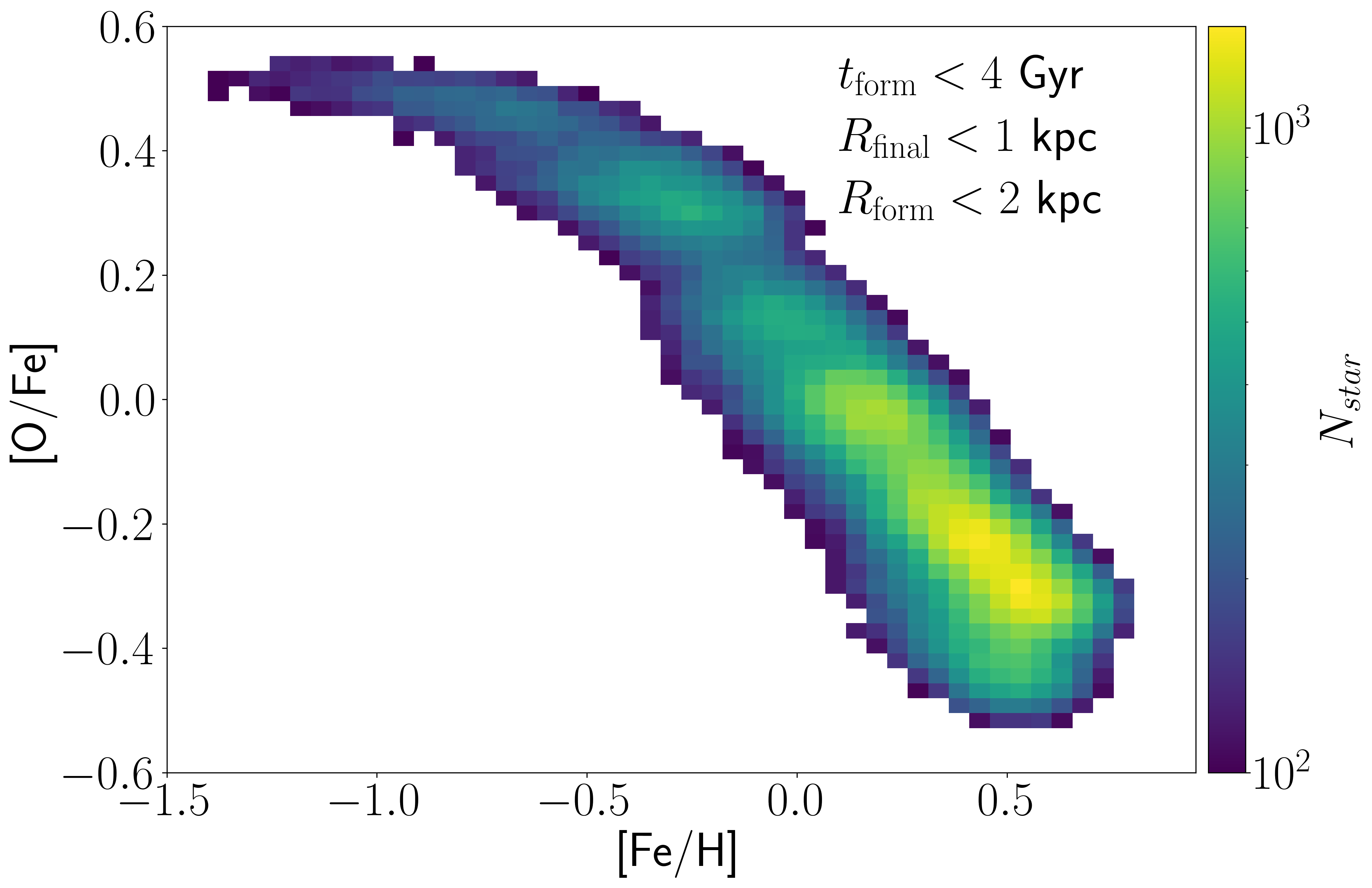}
}
\centerline{
\includegraphics[angle=0.,width=0.5\hsize]{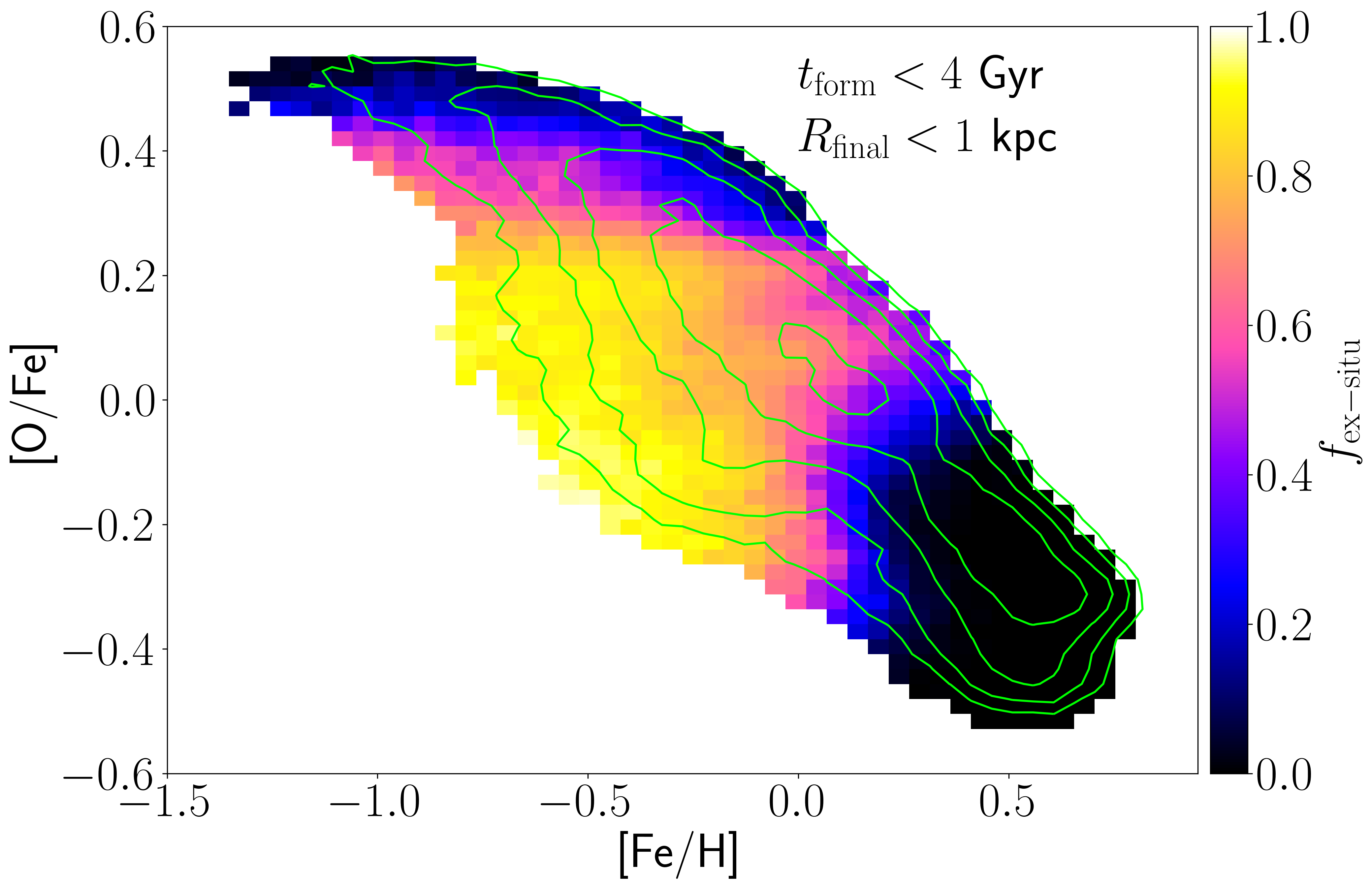}
\includegraphics[angle=0.,width=0.5\hsize]{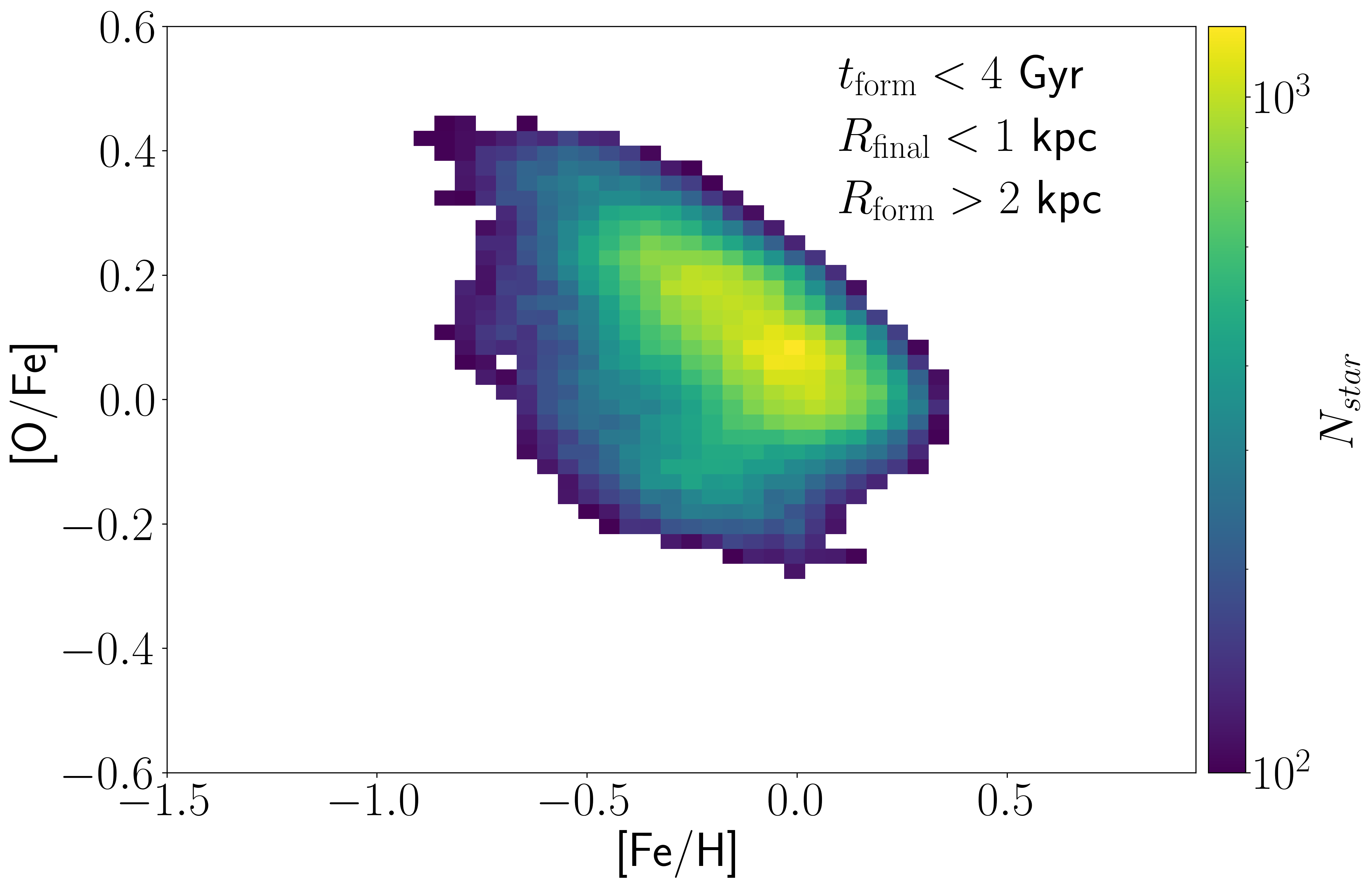}
}
\caption{Top left: distribution of \avg{\rform}\ in the chemical space of stars formed in the first $4~\Gyr$ that end within the inner $1~\kpc$ of the clumpy model. The accreted clumps are responsible for the low-\feh\ peak while in-situ star formation produces the high-\feh\ peak. The contours indicate the density of particles; the 5 contour levels span a factor of 10. 
Bottom left: the fraction of ex-situ stars (those with $\rform > 2~\kpc$) that end up in the bulge ($\rfinal < 1~\kpc$).
Top right: the in situ bulge, showing the distribution of stars contained within $\rfinal < 1~\kpc$ when stars with $\rform > 2~\kpc$ are excluded. While the distribution is not completely smooth, no prominent peak at low-\feh\ is evident.
Bottom right: the ex situ bulge, defined as those stars within $\rfinal < 1~\kpc$ with $\rform > 2~\kpc$. 
\label{f:chemmapRform}}
\end{figure*}

Paper I showed that the clumps in the clumpy simulation often fall to the center.  If clumps are disrupted before they can reach the bulge, then they may play a less prominent role in the formation of the bulge. We therefore consider the formation location of bulge stars to test the effect of the infalling clumps on the chemistry of the bulge.

The top left panel of Fig.~\ref{f:chemmapRform} shows the distribution of the formation radius, \avg{\rform}, in the chemical space. 
An important conclusion from this plot is the different origins of the two MDF peaks. The stars at the low-\feh\ peak in the chemical track have large \avg{\rform}, indicating that many of them are forming outside the bulge and reaching it via clumps. The high-\feh\ peak instead is produced by in-situ\footnote{Here we use the terms in situ and ex situ to refer to formation inside or outside the bulge, but within the galaxy.} star formation (as in the high-feedback model, seen in the top right panel of Fig.~\ref{f:modf}). 
The bottom left panel of Fig.~\ref{f:chemmapRform} shows the fraction of bulge stars that are ex-situ. This is high at intermediate \feh\ and low at high \feh, closely mirroring the top left panel.

When we plot the distribution of stars in the bulge's chemical space excluding those stars that formed outside $\rform = 2~\kpc$, which we do in the top right panel of Fig.~\ref{f:chemmapRform}, we find that the low-\feh\ peak is substantially reduced, with the track resembling somewhat the distribution of the high-feedback model in Fig.~\ref{f:chemicalspace}. (The small peaks remaining after this subtraction are caused by stars formed in clumps that are still star forming inside $R = 2~\kpc$.) This explains why the bulge chemistry at low \feh\ is such a good match to the chemistry of the thick disk: {\it many of these stars share a similar origin}.

The bottom right panel of Fig.~\ref{f:chemmapRform} shows the distribution of those stars excluded from the second panel, \ie\ the stars that end within the bulge that formed at $\rform > 2~\kpc$. This shows that the bulk of these ex-situ stars arriving within clumps settle along the bulge track, with their highest density at the location of the low-metallicity peak.

A number of additional conclusions can be drawn from the top left panel of Fig.~\ref{f:chemmapRform}.  First is the fact that clumps bring with them a small population of low-\al\ stars, which settle below the low-\feh\ peak around $\feh \sim -0.4$ and $\ofe \sim -0.1$. As shown in Paper I (in figures 15 and 17), some of the stars formed in clumps have low \alfe. This population of stars is relatively small and does not contaminate the chemical distribution significantly. The second point is that, to a large extent, the chemistry of the bulge, at the high- and low-metallicity ends, is dominated by stars formed in situ, and is contaminated by clumps only at $-0.5 \la \feh \la 0.0$.

\subsection{The link between the single track and the star formation mode}
\label{ss:sfrd}

\begin{figure*}
\centerline{
\includegraphics[angle=0.,width=0.48\hsize]{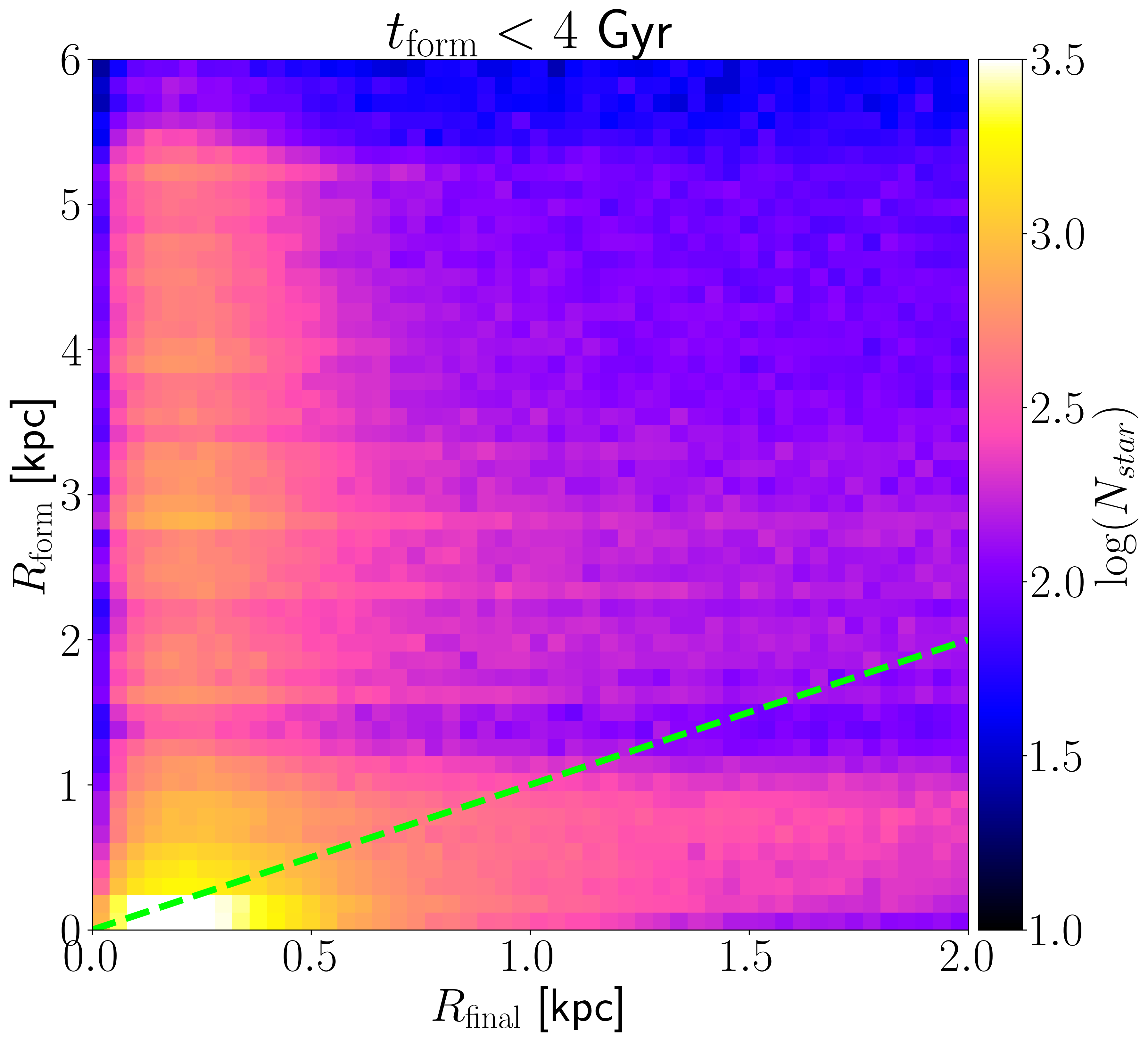}
\includegraphics[angle=0.,width=0.5\hsize]{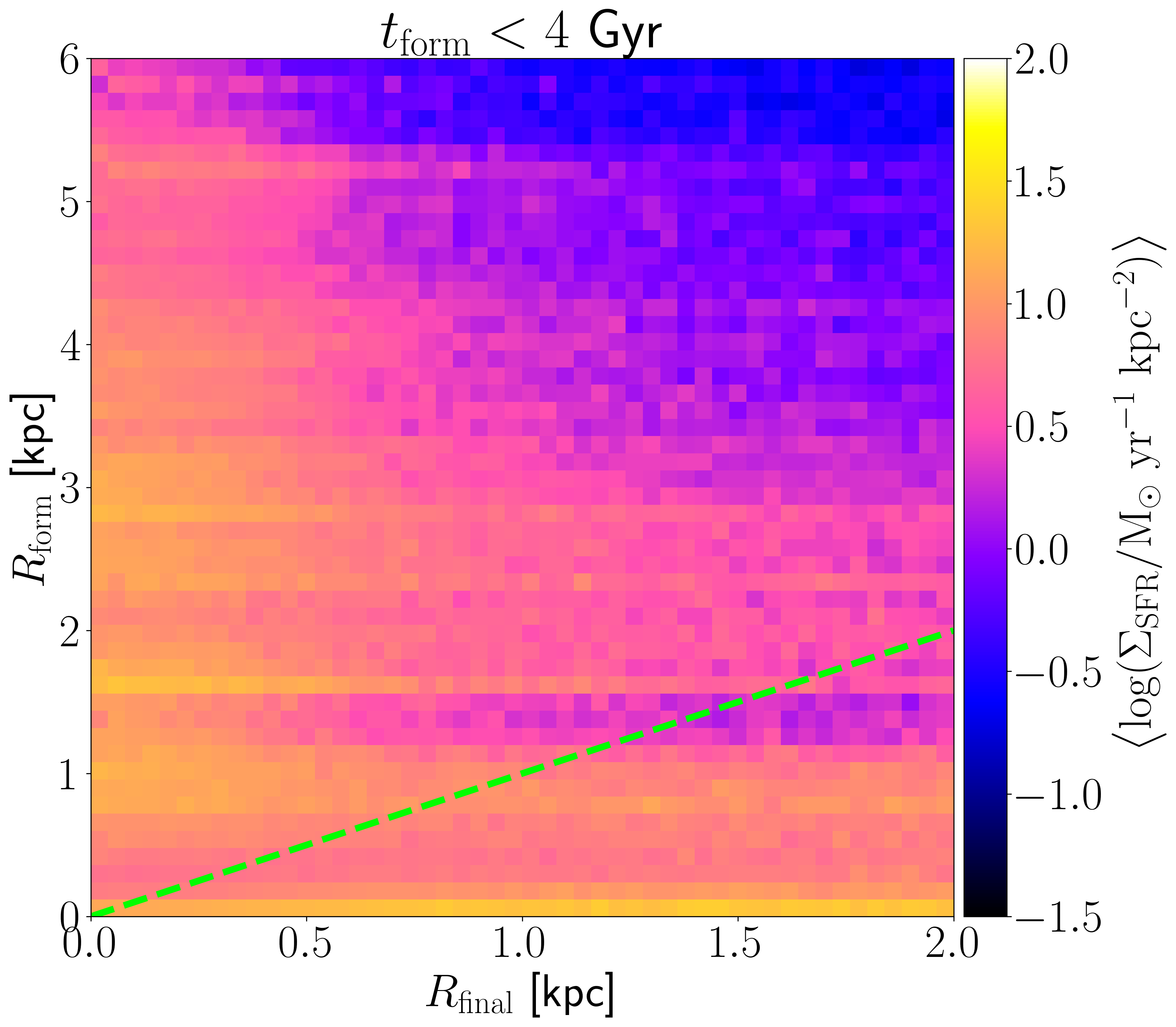}\
}
\centerline{
\includegraphics[angle=0.,width=0.48\hsize]{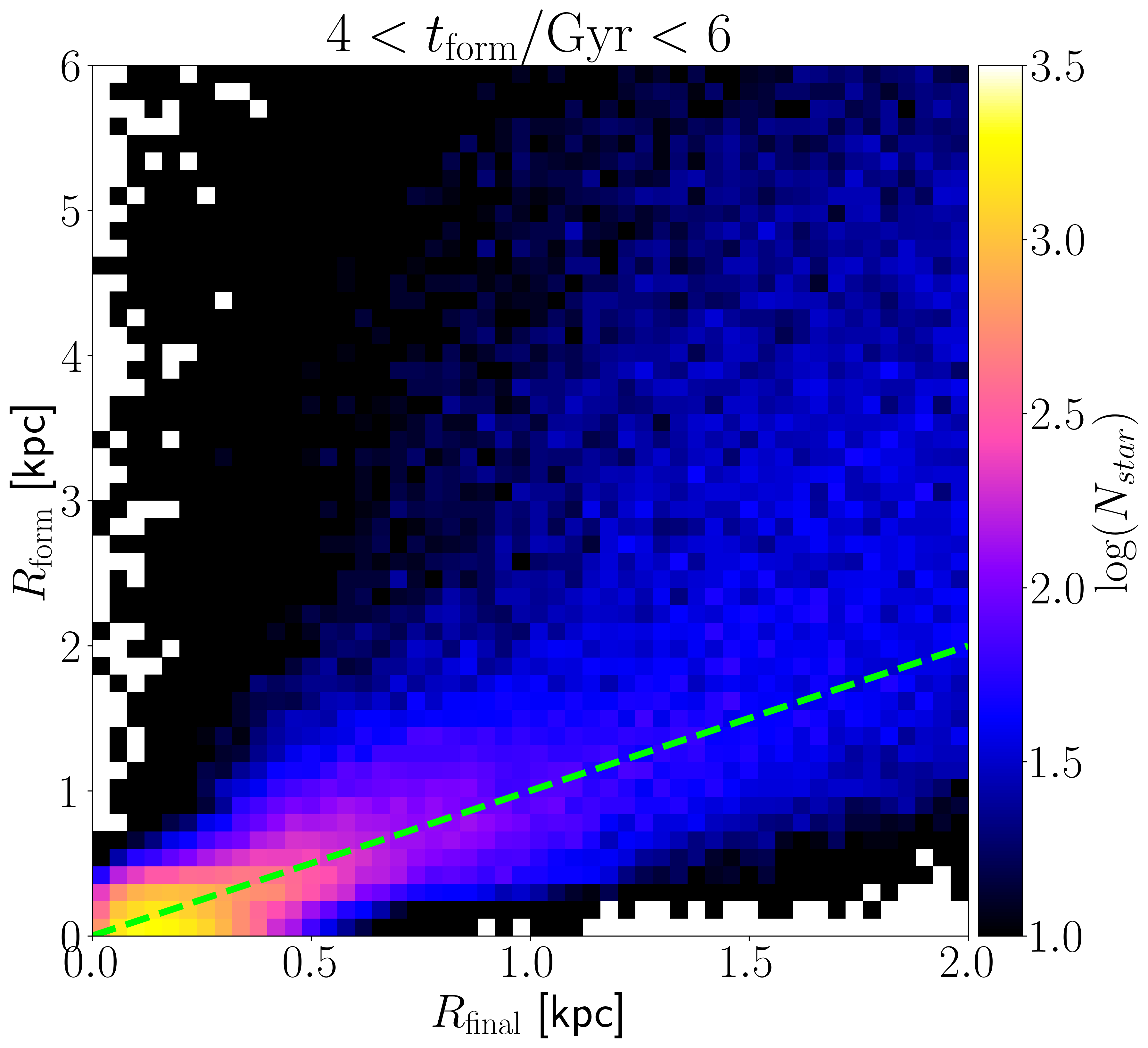}
\includegraphics[angle=0.,width=0.5\hsize]{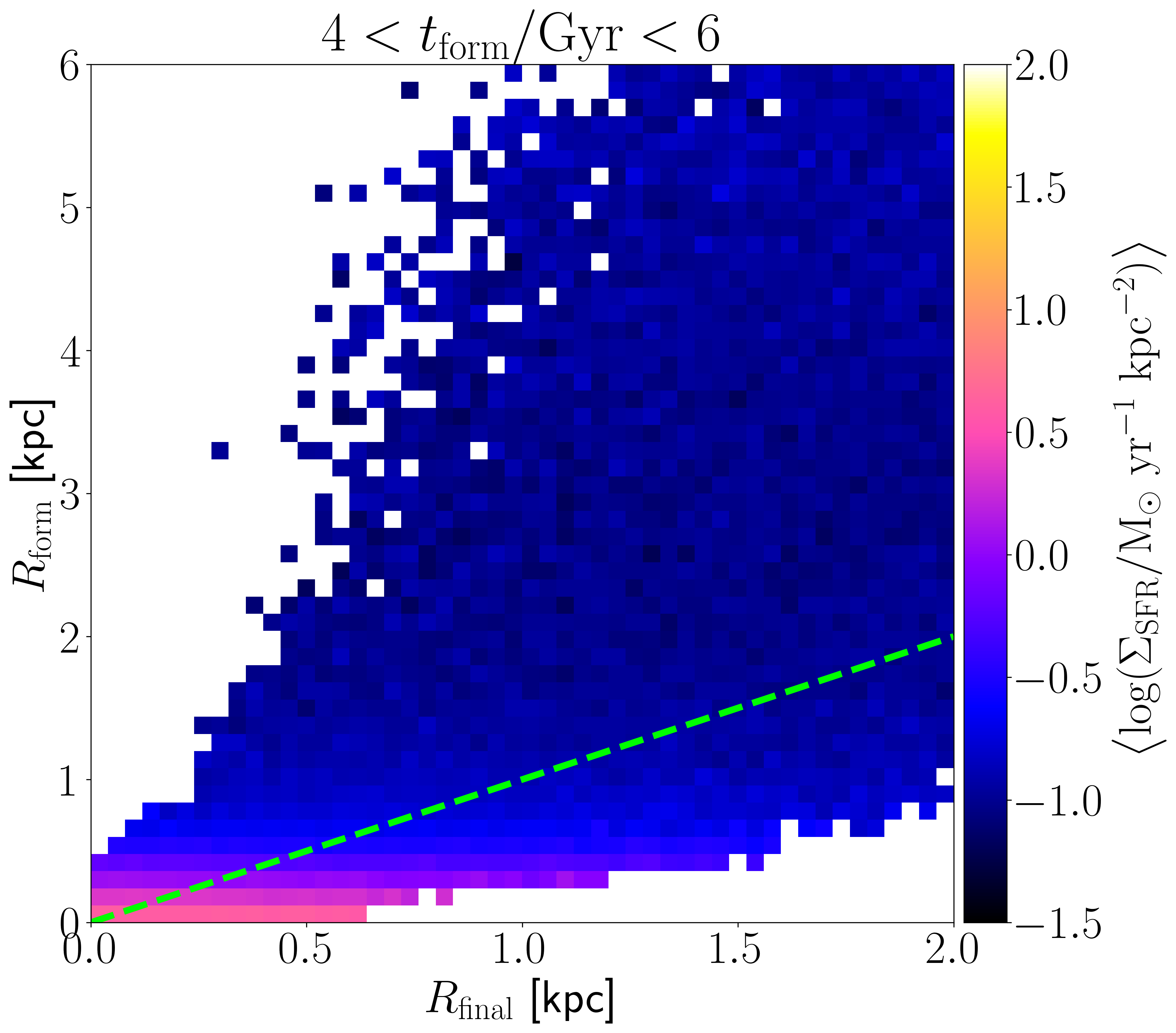}\
}
\caption{The number of stars (left column) and \avg{\sfrd}\ in the \rform-\rfinal\ space for stars at the center of the clumpy model. The top row shows the distributions for $\tform \leq 4~\Gyr$ while the bottom row is for $4 < \tform/\Gyr \leq 6$. The diagonal structure in the bottom panels shows predominantly in-situ star formation after $\tform = 4~\Gyr$, whereas the upper panels show a significant population of bulge stars brought in by clumps. (Note the different scales on the two axes. The diagonal dashed green line indicates $\rform = \rfinal$.)
\label{f:rformrfinal}}
\end{figure*}

In the left panels of Fig.~\ref{f:rformrfinal}, we plot the density of stars in the space of final versus formation radii (\rfinal\ versus \rform). The stars that form during the clump epoch, $\tform < 4\Gyr$, and that end within the inner $1\kpc$ (top left panel) form at a range of radii, including a significant contribution forming in-situ. For the stars formed after the clump epoch, $4 \leq \tform/\Gyr \leq 6$, (bottom left panel) star formation occurs in situ, resulting in the diagonal distribution in the \rfinal-\rform\ space. In the absence of clumps, stars only reach the bulge from farther out via eccentric orbits; the bottom left panel shows that the fraction of such stars is low. 

The stars in the bulge therefore are a mix of those formed in situ and those accreted in clumps. Paper I showed that there are two modes of star formation: a high \sfrd\ and a low \sfrd\ one. Clumps are associated with high \sfrd\ ($\sfrd \ga 1~\Msun~\yr^{-1}~\kpc^{-2}$) while lower \sfrd\ is typical of distributed (nonclumpy) star formation (see Figure 15 of Paper I).
The right panels of Fig.~\ref{f:rformrfinal} show the distribution of \avg{\sfrd}\ in the same \rfinal\ versus \rform\ space. For stars with $\tform < 4~\Gyr$ (top right panel), the high \avg{\sfrd}\ at $\rfinal < 0.5~\kpc$ is produced by the full range of \rform, which therefore must include stars formed in clumps that have fallen in, as well as those formed in situ. This is made clearer by comparing with stars that form after $4~\Gyr$ (bottom right panel) when clump formation has ceased; now bulge stars have relatively low $\avg{\sfrd} \simeq 3~\Msun~\yr^{-1}~\kpc^{-2}$ (except at the very center), and $\rform \simeq \rfinal$. These stars clearly are forming in situ rather than falling in as clumps.

\begin{figure}
\includegraphics[angle=0.,width=0.6\hsize]{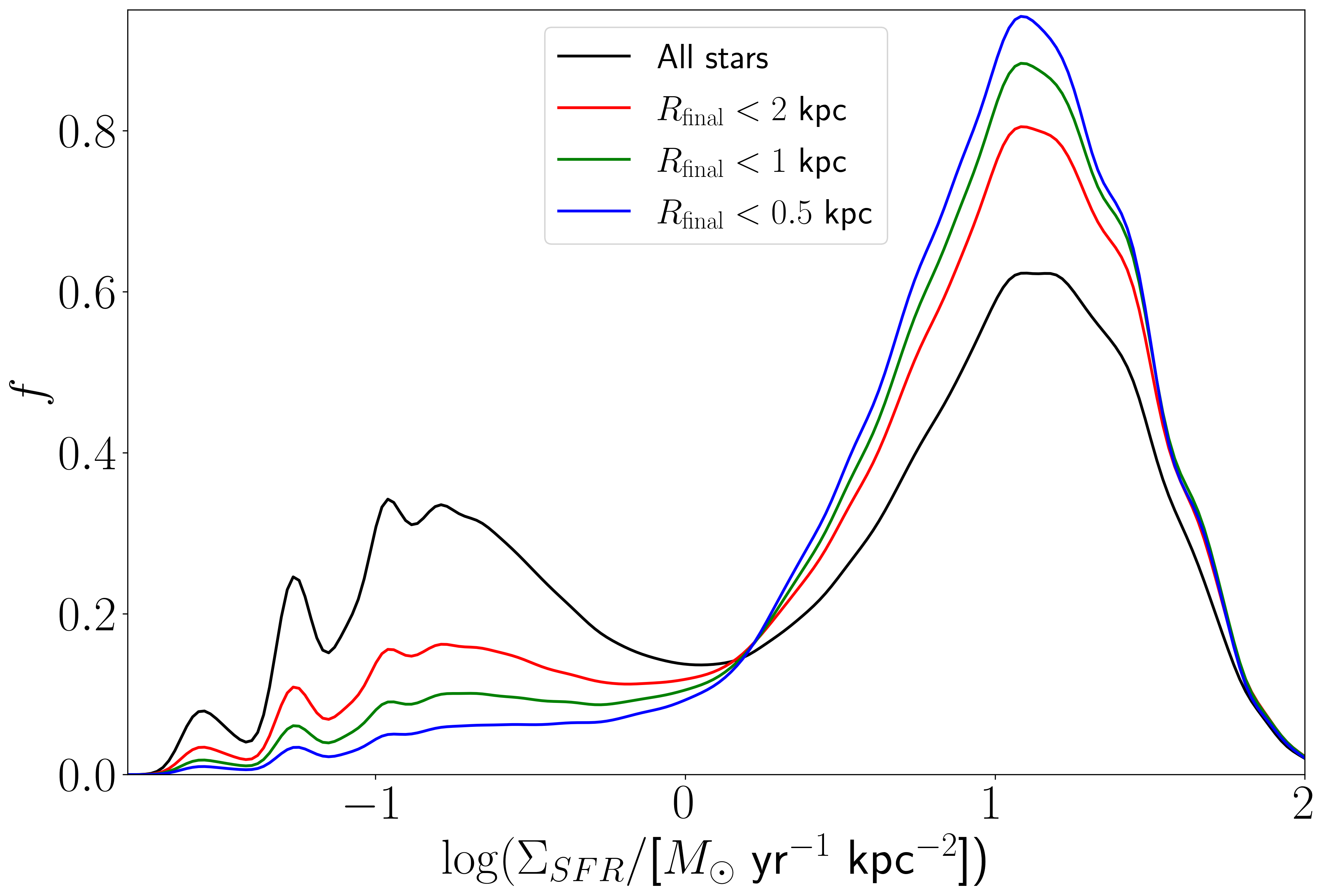}\\
\includegraphics[angle=0.,width=0.6\hsize]{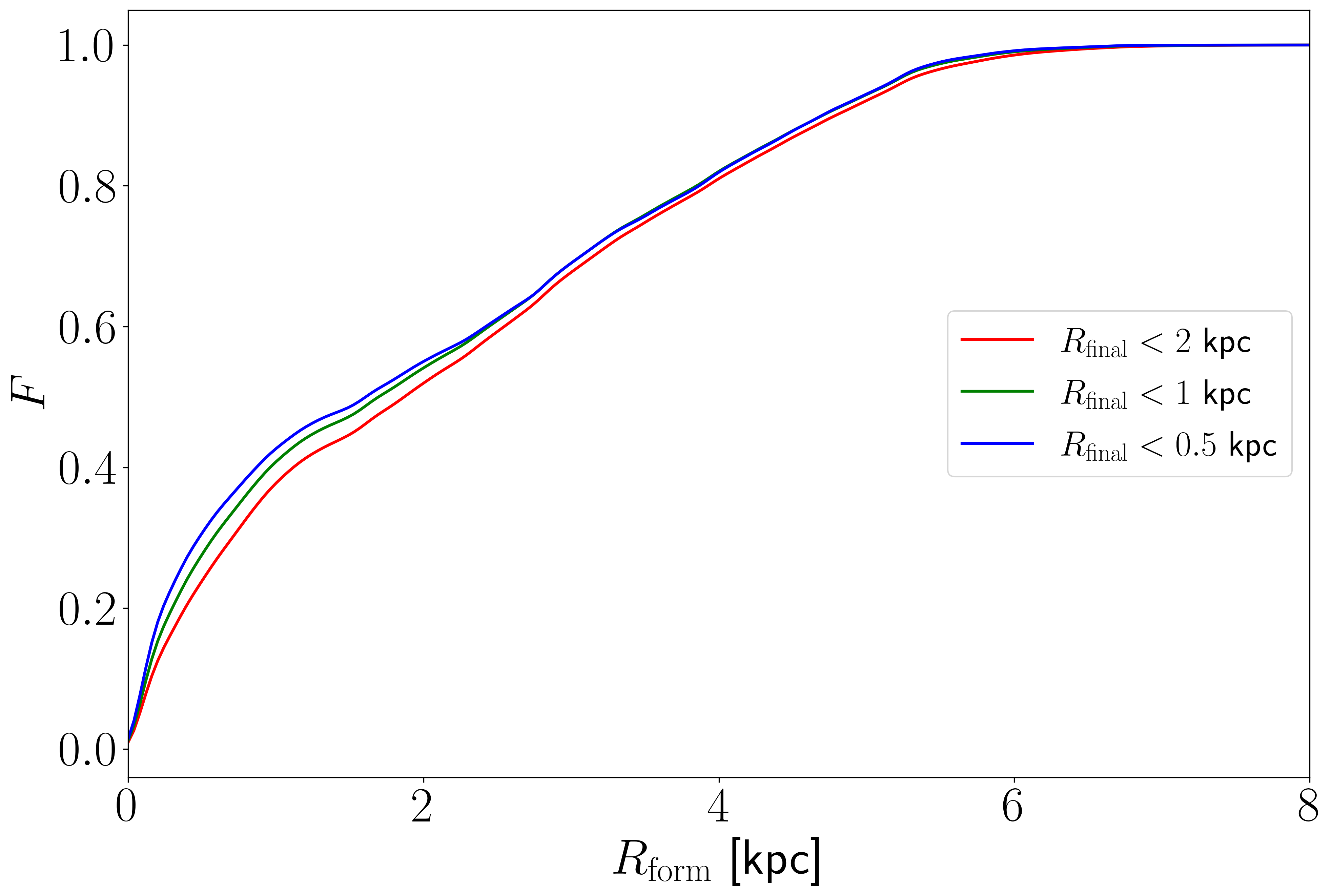}\\
\includegraphics[angle=0.,width=0.6\hsize]{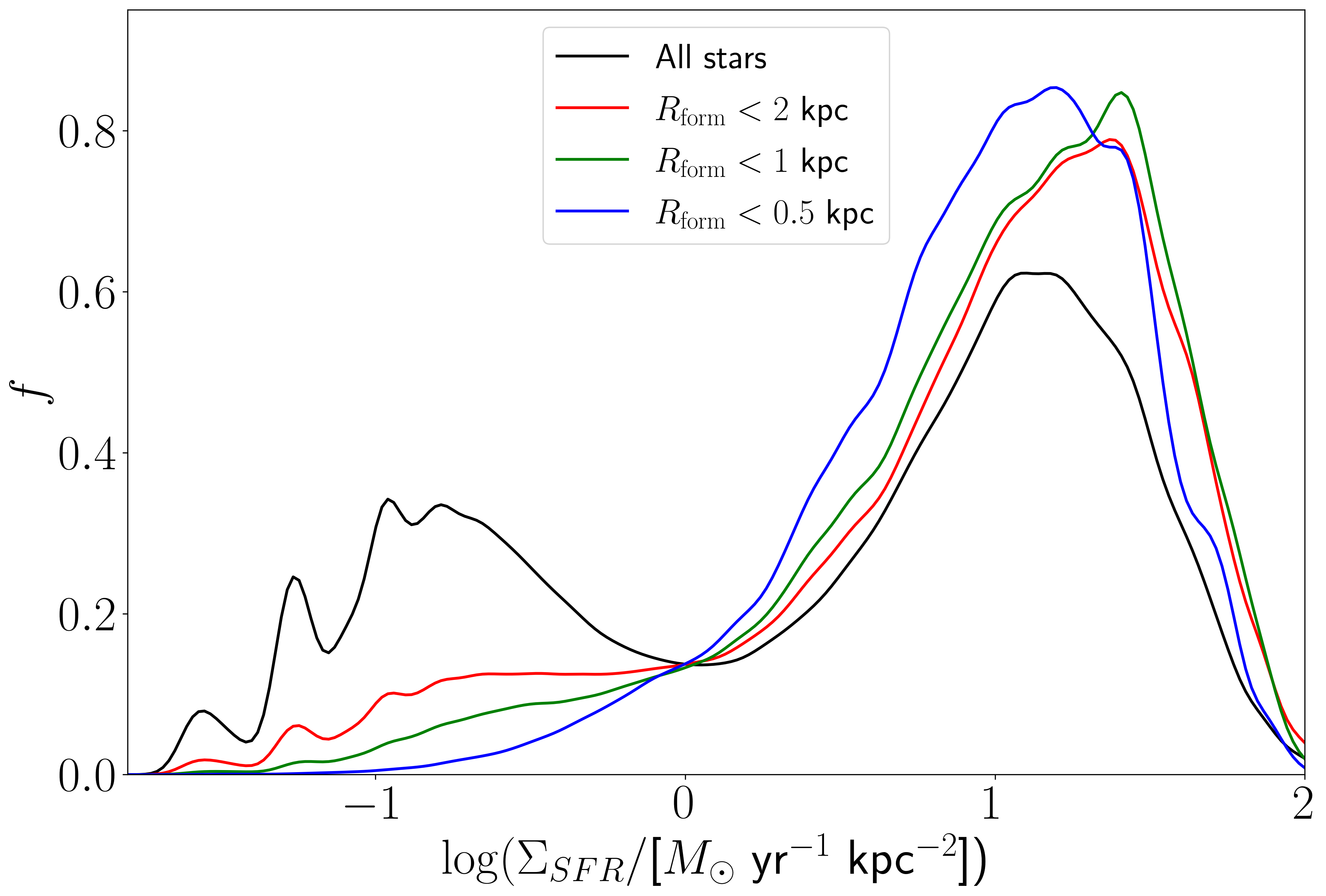}\
\caption{Star formation modes at the center of the clumpy model for stars with $\tform \leq 4 \Gyr$. Top: The normalized distribution of the star formation rate density for different final radii.  Middle: The cumulative formation radius of stars that end up at different radii.  Bottom: The normalized distribution of star formation rate density for different formation radii. In the top and bottom panels, the black histograms refer to all the stars within the model. All distributions use kernel density estimates (KDEs) with a Gaussian kernel and window width satisfying Silverman's rule \citep{silverman86}.
\label{f:formation}}
\end{figure}

Therefore stars born outside the bulge in the clumpy high-\sfrd\ mode are reaching the bulge. In the top panel of Fig. \ref{f:formation} we plot \sfrd\ for stars forming before $4\Gyr$ that end at different radii within the inner galaxy. From $\rfinal \leq 2~\kpc$ (red curve) to $\rfinal \leq 1~\kpc$ (green curve), the contribution of the high-\sfrd\ mode of star formation rises, and overwhelmingly dominates at $\rfinal < 0.5~\kpc$ (blue curve). This would seem to imply that infalling clumps dominate the inner galaxy. However, the middle panel of Fig.~\ref{f:formation}, which shows the cumulative distribution of $\rform$ for stars at different \rfinal, 
shows that a significant in-situ population is also present. Indeed the fraction of stars that formed within $\rform = 2~\kpc$ {\it rises} as \rfinal\ decreases, although it never exceeds $\sim 60\%$.  Roughly half the stars that end up at $\rfinal \leq 2~\kpc$ were born outside this region. Thus clumps are delivering a significant fraction of the bulge's mass, but in-situ star formation is equally important.

Why then does the in-situ star formation not produce a separate track in the bulge chemistry like the disk's low-\alfe\ track? The bottom panel of Fig.~\ref{f:formation} plots \avg{\sfrd}\ for stars that were born within a given radius by $\tform = 4\Gyr$.  The vast majority of early stars formed within $2~\kpc$ formed via the high-\sfrd\ mode. Therefore, the early bulge itself acts as a clump of high \sfrd, as first shown by \citet{mandelker+14}. Thus the bulge never gets to form a low-\al\ track: even in the absence of clumps falling into the bulge, for instance because they are disrupted before they reach the center, the bulge chemistry will still lack metal-poor low-\al\ stars. Indeed even the high-feedback model has only a single track in the bulge, and is \al-rich (compared with the disk), as can be seen in the top right panel of Fig.~\ref{f:chemicalspace}.

\subsection{Bulge ages and quenching}
\label{ss:ages}

\begin{figure}
\includegraphics[angle=0.,width=0.6\hsize]{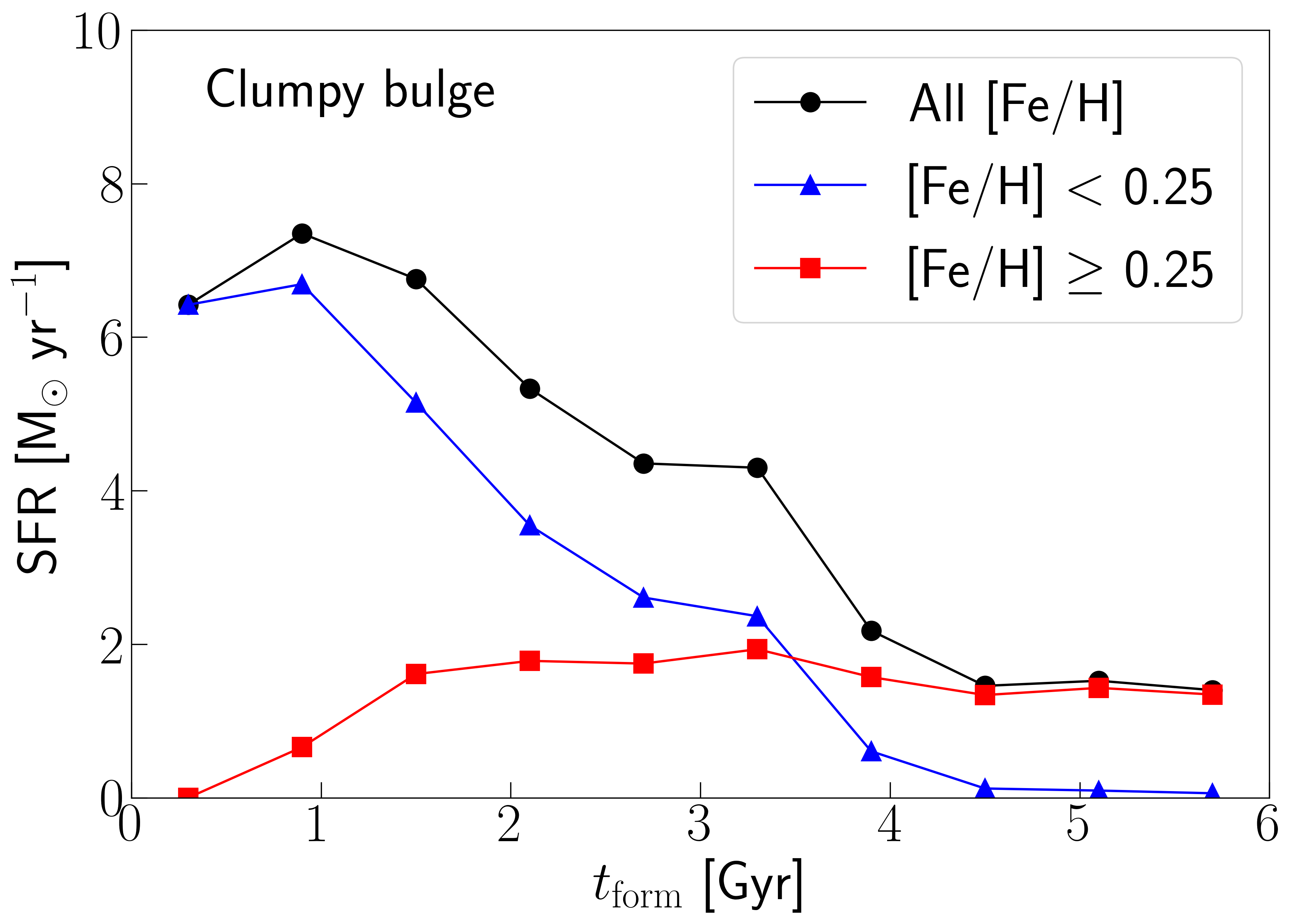}\\
\includegraphics[angle=0.,width=0.6\hsize]{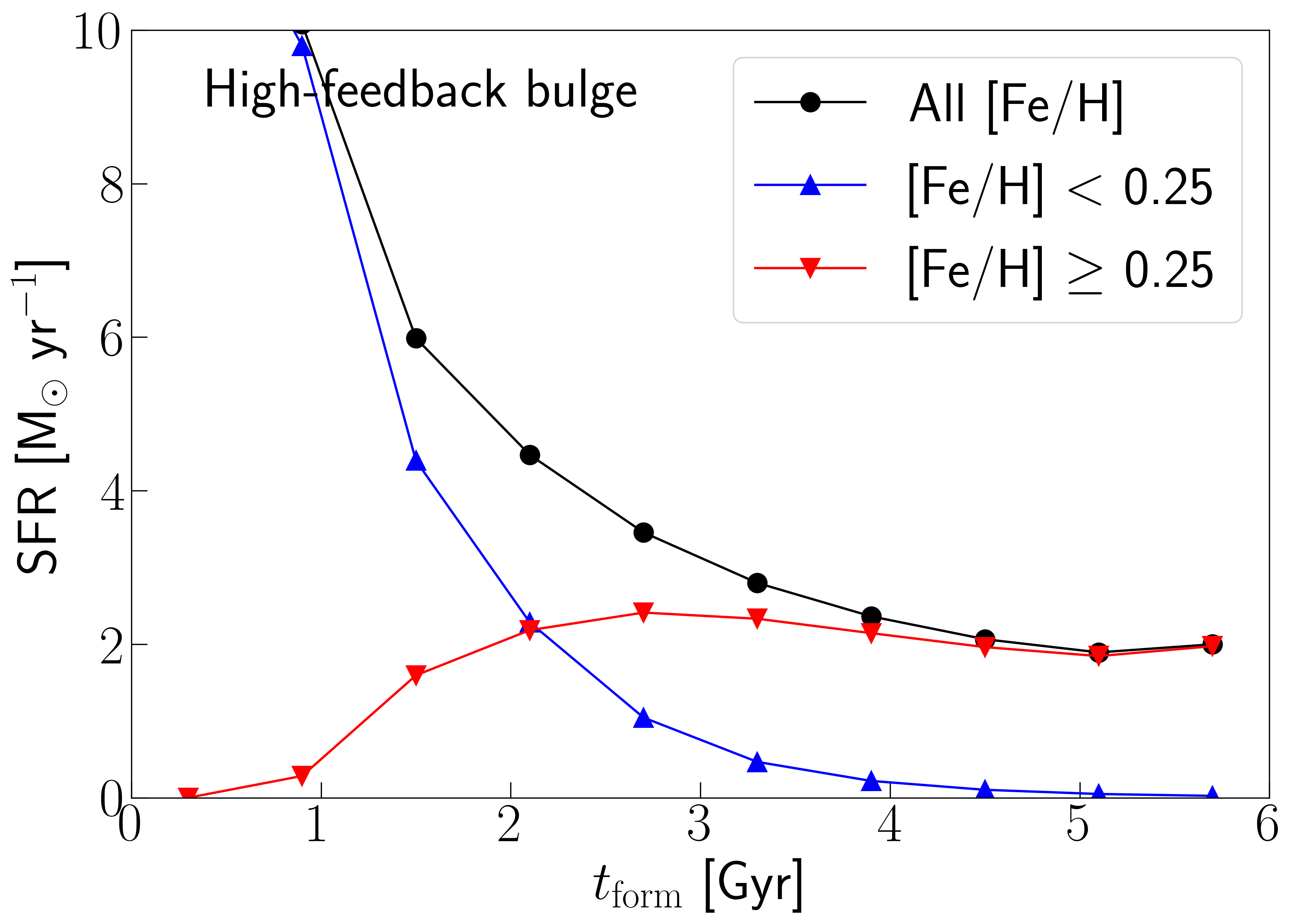}
\caption{Star formation history to $6\Gyr$, \ie, $2\Gyr$ after the end of the clump era in the clumpy model, for stars that end up within $\rfinal < 1\kpc$, separated by metallicity. The full distribution is shown in black. The separation into metal-rich and metal-poor is at $\feh= 0.25$, which marks the minimum between the two peaks in the MDF of the clumpy model. The stars in the low-\feh\ and high-\feh\ populations are shown in blue and red, respectively. The top panel shows the clumpy model while the bottom panel shows the high-feedback model split at the same metallicity. The overall similarity of the SFHs of the two models suggests that quenching is not responsible for the trough in the chemical space of the clumpy model, seen in the top left panel of Fig.~\ref{f:chemicalspace}.
\label{f:adf}}
\end{figure}

If star formation continues in the bulge after the clump epoch ends, then the younger stars will necessarily be at the high-\feh\ peak. The high-\feh\ peak then would be younger, on average, than the low-\feh\ peak. In the MW, the difference in mean age between the two peaks would be governed by when the bar forms, because bars generally quench star formation within most of their radius (including the vertically thickened part that forms the bulge). Fig.~\ref{f:adf} shows the SFH up to $6~\Gyr$ for the stars that end within $R = 1~\kpc$ (we have checked that the result does not change qualitatively if we consider stars inside $R = 2\kpc$). This shows that the high-metallicity population ($\feh \geq 0.25$) overlaps in age with the low-\feh\ population. However no new stars with low-\feh\ form after $\sim 4\Gyr$. 

\citet{lian+20} interpreted the trough between the two peaks in the MW bulge's chemical track as an episode of quenching in its SFH, before star formation restarted and produced the high-\feh\ peak. The top panel of Fig.~\ref{f:adf} shows that the star formation in the clumpy model's bulge never drops to zero, although its chemical track in Fig.~\ref{f:chemicalspace} develops a trough between the two peaks. The bottom panel of Fig.~\ref{f:adf} presents the SFH of the high-feedback model. Despite the similarity in the SFH of the two models, only the clumpy model develops a trough between two peaks in the bulge chemical track (as can be seen in Figure~\ref{f:chemicalspace}), suggesting that the SFH need not be responsible for the trough.

\begin{figure}
\includegraphics[angle=0.,width=\hsize]{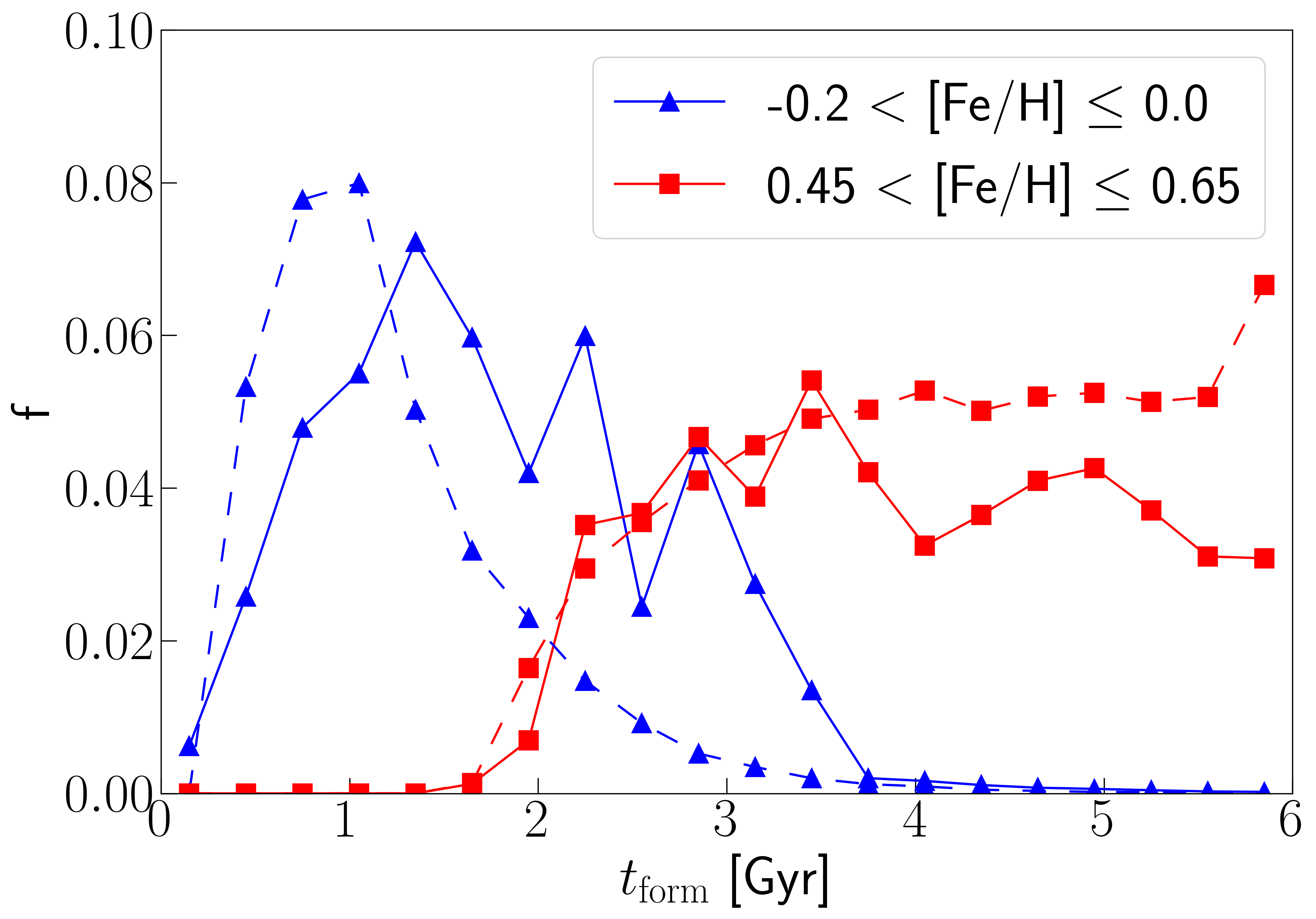}
\caption{The SFH of stars centered at the clumpy model's two MDF peaks within the inner $1~\kpc$. The clumpy model is shown by the solid lines while the high-feedback model at the same \feh\ ranges is shown by the dashed lines. The two metallicity ranges selected are indicated by the vertical dotted lines in the top row of Fig.~\ref{f:modf}.
\label{f:adfinpeaks}}
\end{figure}

The evolution of the bulge MDF in the clumpy model, seen in Fig.~\ref{f:modf}, shows that the bulge reaches the high-\feh\ regime already by $2\Gyr$, and is bimodal already by that point. The bimodality increases at later times, particularly after $4\Gyr$, but the trough is present before the clumpy episode is over. Thus the high-\feh\ peak contains old stars and represents the ordinary chemical evolution of a rapidly star-forming system. If we understand the \feh-enrichment as developing smoothly, then stars in the trough will be, on average, slightly older than those in the high-\feh\ peak. The stars at the low-\feh\ peak, because they are a mix of in-situ stars and stars accreted via clumps, represent a range of ages, from older than the trough (from the in-situ evolution) to stars younger than the old stars in the high-\feh\ peak (from the later stages of clump accretion). Fig.~\ref{f:adfinpeaks} shows the distribution of ages at the two peaks, within the \feh\ limits indicated by the vertical dotted lines in the MDFs of Fig.~\ref{f:modf}. At $2 \lesssim \tform/\Gyr \lesssim 4$, the ages of stars at the high-\feh\ peak significantly overlap those of the youngest stars at the low-\feh\ peak. We show, as dashed lines, the age distributions for the high-feedback model in the same metallicity ranges. While the age distribution at the high-\feh\ peak is comparable to that of the clumpy model, the region where the low-\feh\ peak would be has predominantly older stars and only overlaps the high-\feh\ peak's ages in the exponential wing of the distribution.

\begin{figure}
\includegraphics[angle=0.,width=0.9\hsize]{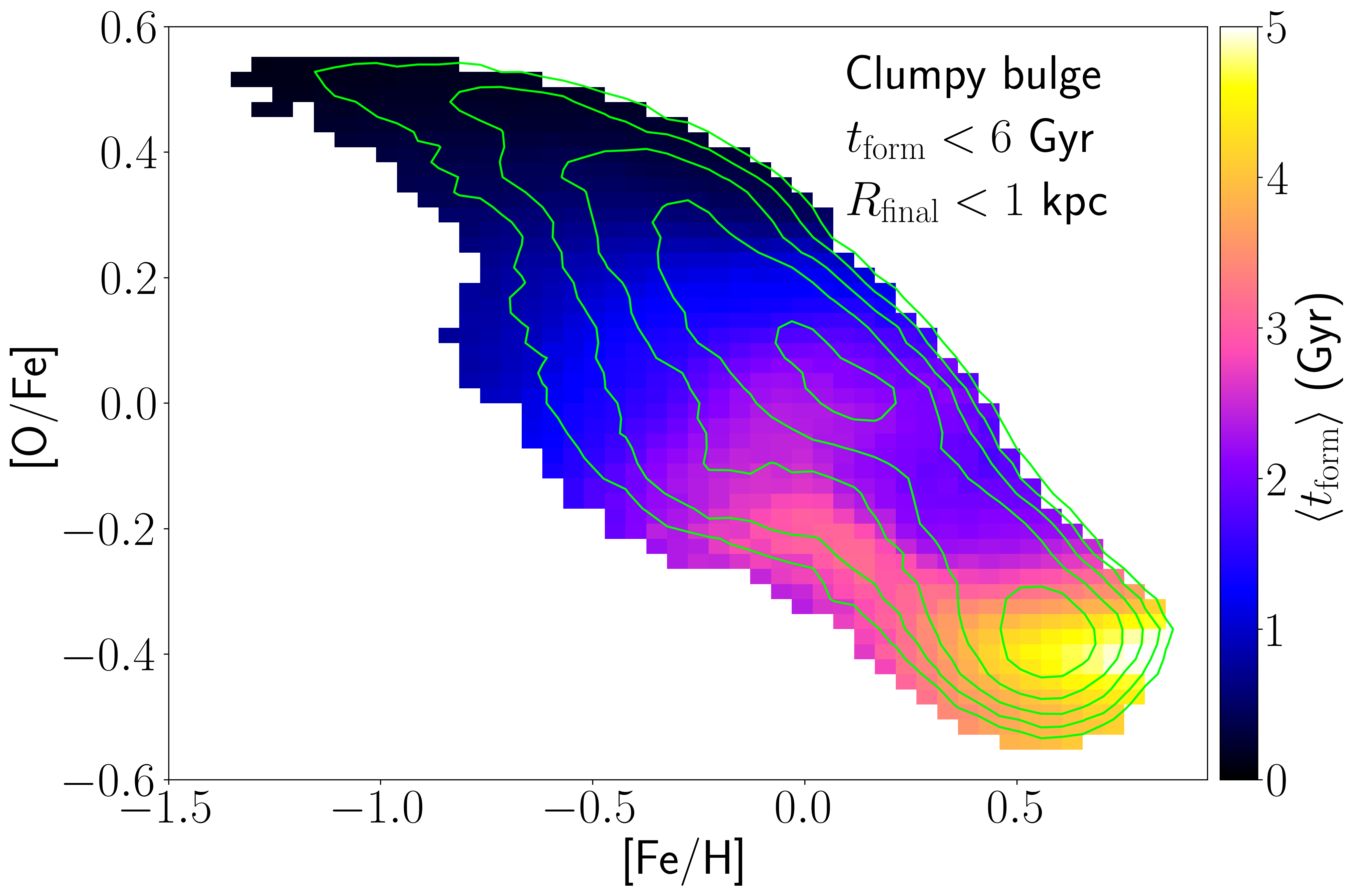}
\includegraphics[angle=0.,width=0.9\hsize]{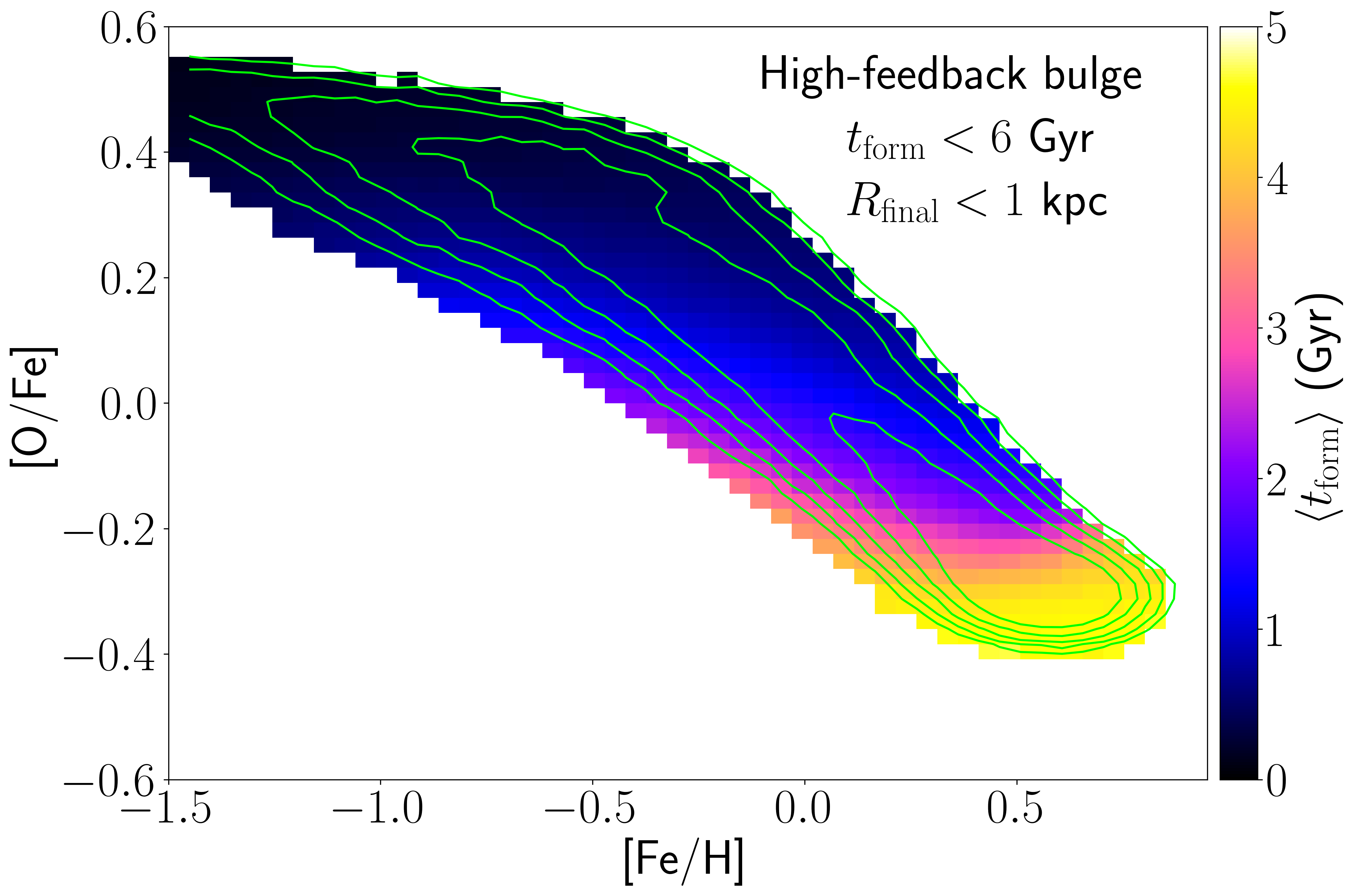}
\caption{The distribution of \avg{\tform} in the chemical space of stars formed in the first $6~\Gyr$ that end within the inner $1~\kpc$ of the clumpy (top) and high-feedback (bottom) models. The location of the trough in the chemical space of the clumpy model corresponds to a local minimum in \avg{\tform}. The contours indicate the density of particles; the 5 contour levels span a factor of 10. 
\label{f:quenching}}
\end{figure}

We conclude that the trough between the high- and low-\feh\ peaks in the clumpy model is not due to a quenching of in-situ star formation but rather due to the end of clumps delivering stars to the low-\feh\ peak of the bulge. A possible diagnostic of this scenario is the age distribution of the low-\feh\ peak compared with that in the trough: stars at the low-\feh\ peak should include younger  stars than those in the trough. We explore this prediction for the clumpy model in the top panel of Fig.~\ref{f:quenching}, where we plot the mean time of formation, \avg{\tform}, of stars in the chemical space of the bulge stars formed by $t=6~\Gyr$. {\it Along the ridge} of the chemical track from metal-poor to metal-rich, we reach a local maximum in \avg{\tform} at the location of the low-\feh\ peak, while the high-\feh\ peak is the location of late star formation and has the largest \avg{\tform} (\ie, youngest stars). In between, at the trough, \avg{\tform} has a local minimum, meaning the stars in this region are older. Observationally, this dip gives the appearance of a drop in the SFR of the bulge and thus resembles a quenching episode. However this is clearly not the case in the evolution of the clumpy model. In contrast, the bottom panel shows the mean age of the high-feedback model, which shows that the mean age increases monotonically along the ridge in this case.

A final noteworthy property of the clumpy bulge's chemistry is that the trough occurs just beyond the highest metallicity of the thick disk track (see Fig.~\ref{f:chemicalspace}). This happens because the trough is not significantly polluted by stars formed in the same clumps that produced the thick disk.

\subsection{Dependence of kinematics on chemistry}
\label{ss:alphakine}

\begin{figure}
\includegraphics[angle=0.,width=0.5\hsize]{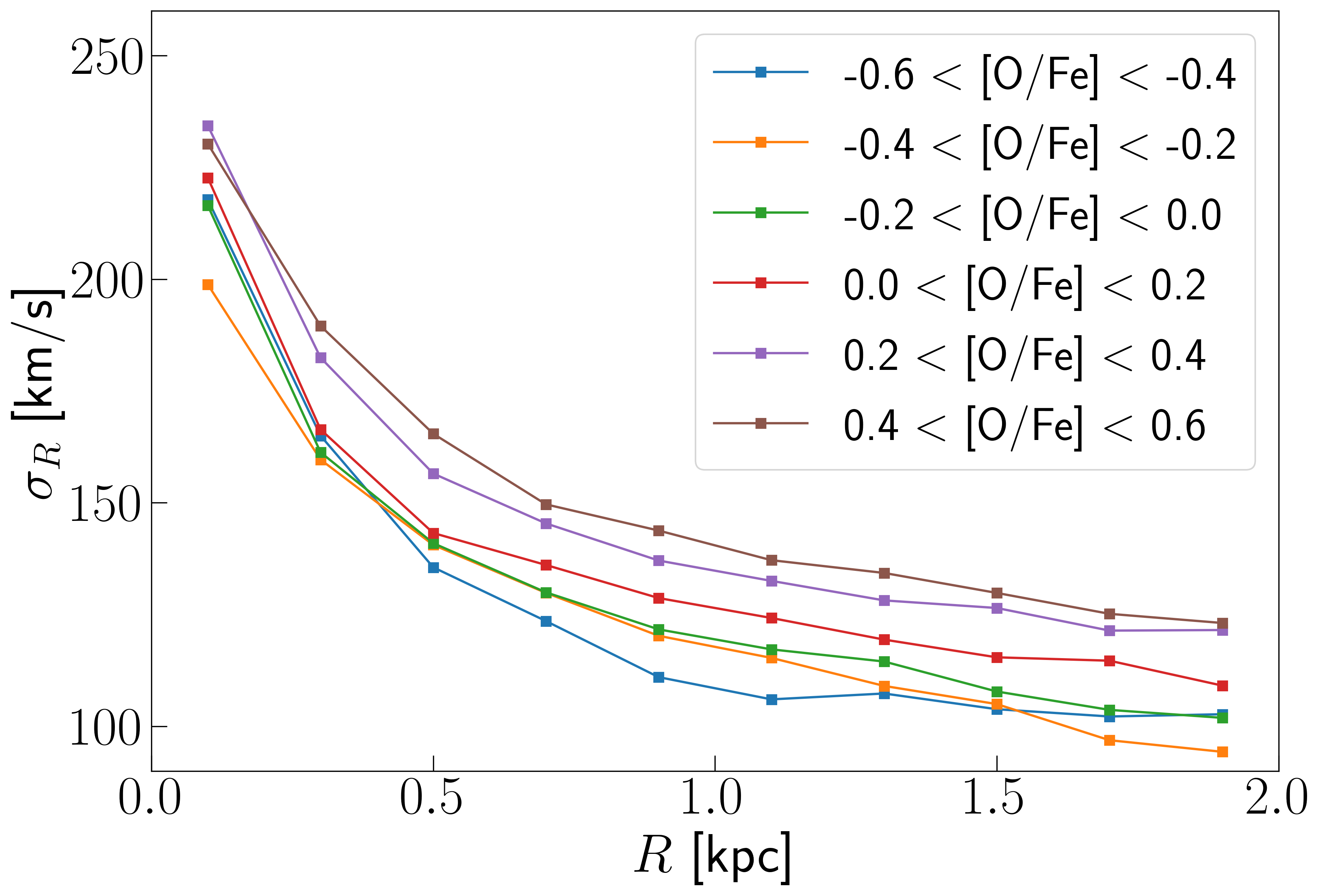}\\
\includegraphics[angle=0.,width=0.5\hsize]{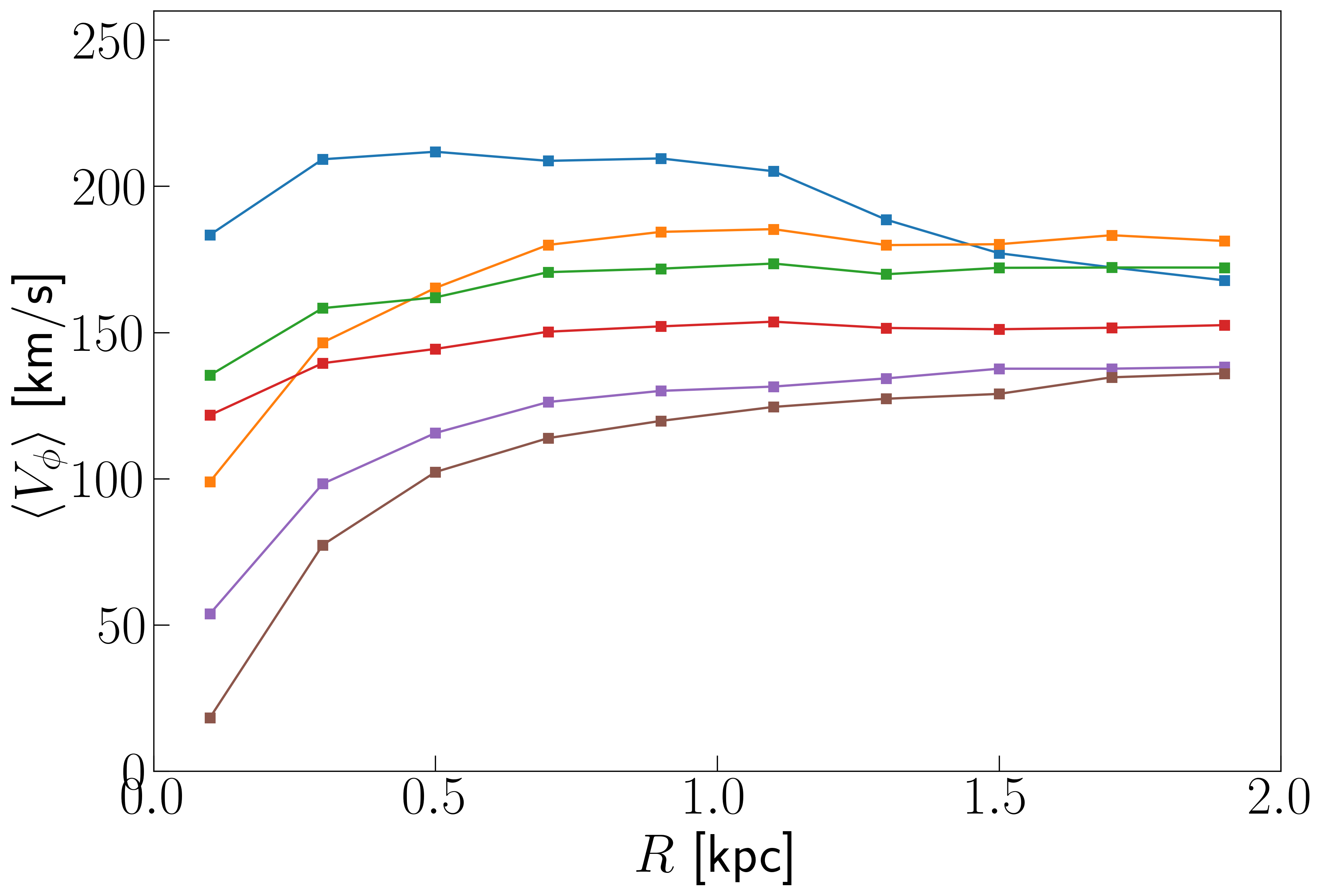}
\caption{Radial velocity dispersions, \sig{R}, (top) and mean rotational velocity, \avg{V_\phi}, (bottom) of stars at $4\Gyr$, in the inner $2\kpc$ of the clumpy model, as a function of \ofe. Low-\ofe\ stars have lower radial velocity dispersions and higher rotation.
\label{f:sigR}}
\end{figure}

\begin{figure}
\includegraphics[angle=0.,width=\hsize]{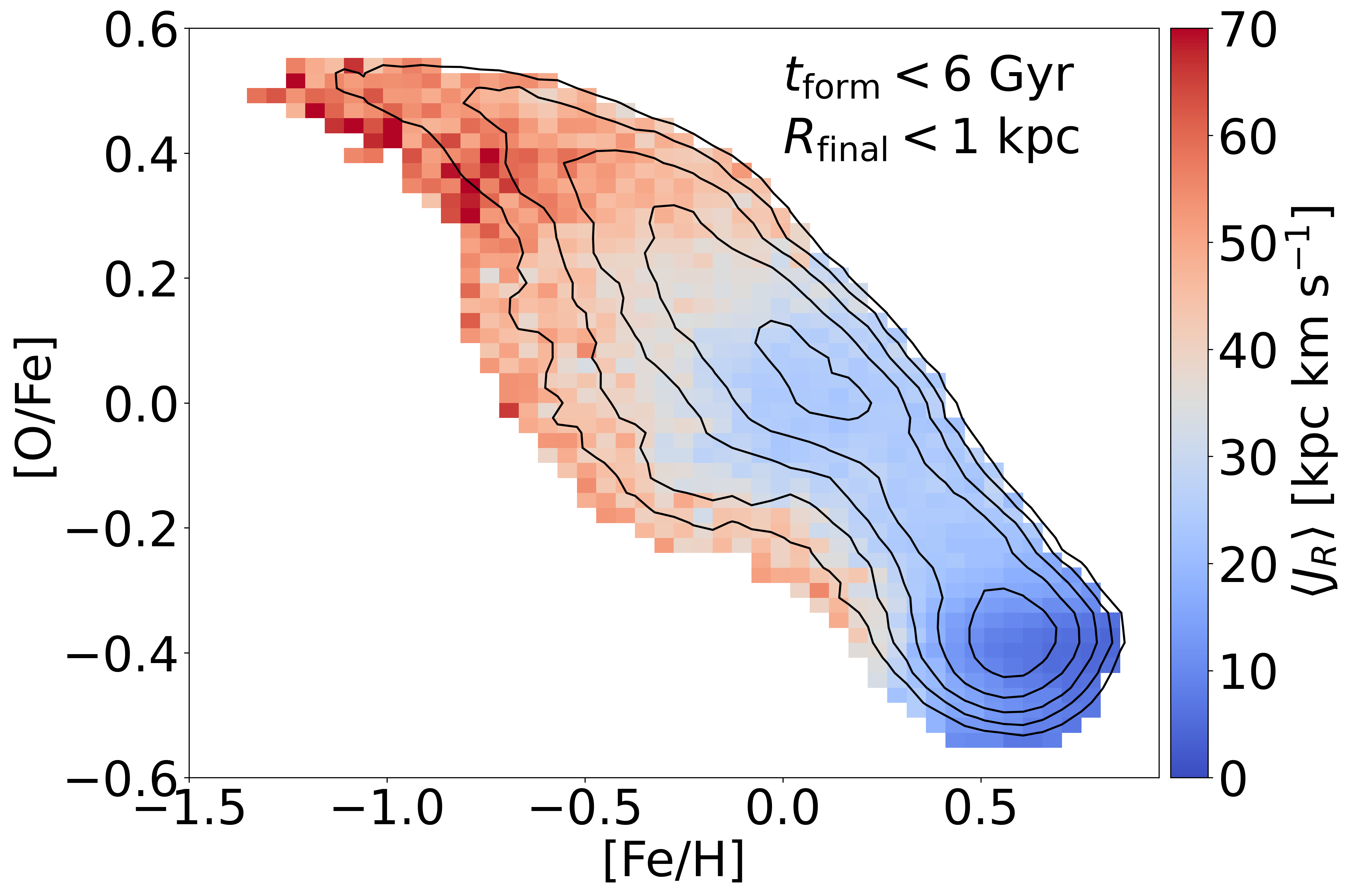}
\caption{The mean radial action, \avg{J_R}, in the chemical space of the clumpy model. The contours indicate the density of particles; the 5 contour levels span a factor of 10.
\label{f:chemdispmap}}
\end{figure}

The top panel of Fig.~\ref{f:sigR} shows profiles of the radial velocity dispersion, \sig{R}, of stars at $t=4\Gyr$ separated into \ofe\ bins. The high-\ofe\ stars are generally hotter, by $20-30~\kms$, than the low-\ofe\ stars even just at the end of the clumpy epoch. This reflects on the chaotic interaction of clumps near the center of the galaxy, which heats the high-\ofe\ populations at birth. 

The bottom panel of Fig.~\ref{f:sigR} shows profiles of the mean streaming velocity, \avg{V_\phi}; although clumps are falling to the center, the bulge remains rotationally supported, because the clumps are on in-plane, prograde orbits, which are known to produce rapidly rotating remnants even when the resulting mergers are collisionless \citep[e.g.][]{read+08, hartmann+11}. In cosmological simulations, \citet{inoue_saitoh12} also found rapidly rotating bulges forming from clumps. The low-\ofe\ stars are more rapidly rotating, by $\sim 50-100\kms$, as expected given their lower velocity dispersion.

We measure the actions of stars using {\sc agama} \citep{agama}, which uses the St\"ackel fudge of \citet{binney12}, assuming a flattened axisymmetric potential for the disk and a spherical potential for the halo.
Fig.~\ref{f:chemdispmap} shows the mean radial action, \avg{J_R}, in the chemical space, for stars in  the  inner  $1~\kpc$ at $6~\Gyr$. Bearing in mind that \citet{debattista+20} found that bar formation substantially steepens the vertical gradient of \avg{J_R}, we anticipate that stars at the high-\feh\ peak would dominate near the mid-plane while the large heights would be dominated by the low-\feh\ peak if a bar had formed.

\subsection{Comparison with the Milky Way}
\label{ss:mwcomp}

\begin{figure}
\includegraphics[angle=0.,width=0.8\hsize]{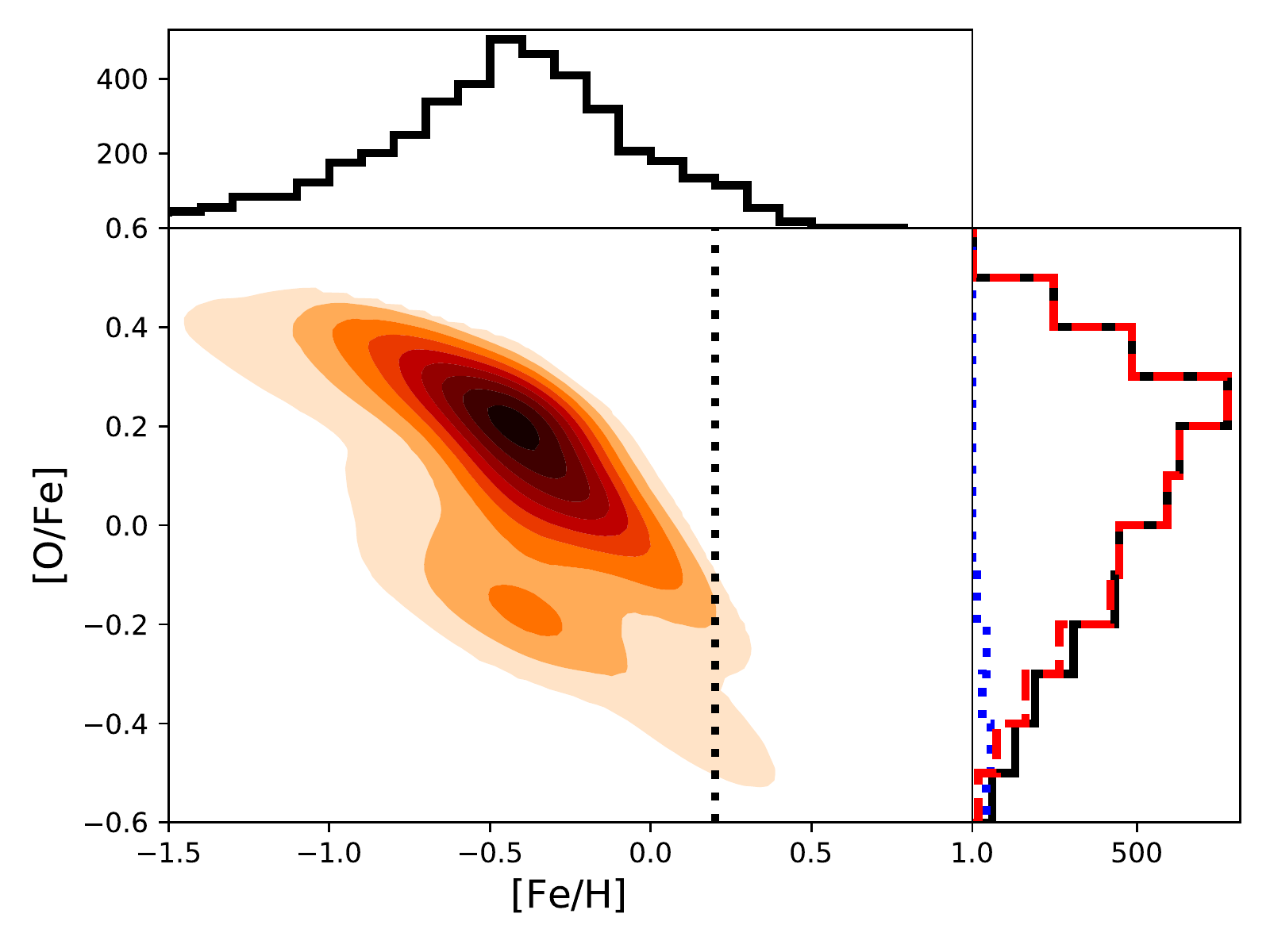}
\includegraphics[angle=0.,width=0.8\hsize]{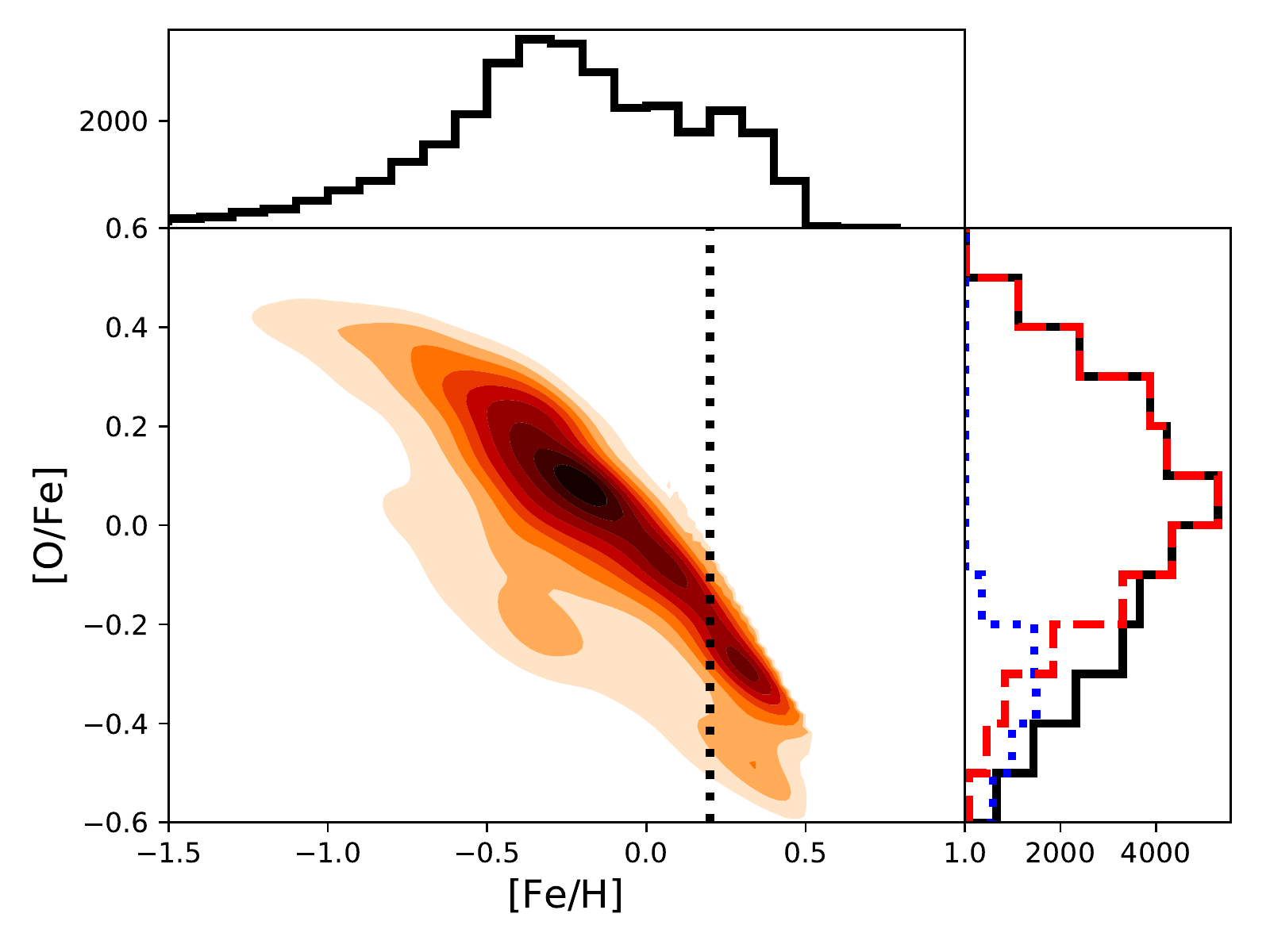}
\caption{Gaussian kernel estimate of the density in the \feh-\ofe\ chemical space for stars in the clumpy model. Ten equally spaced contours show the density distribution of stars selected to be within the volume bounded by $-6.5\degrees < l < 6.5\degrees$ and $7<d/\kpc<9$, in $1\degrees$ stripes at $b=2\degrees$ (bottom) and $6\degrees$ (top). The histogram at the top shows the full distribution of \feh; the minimum between the two \feh\ peaks, indicated by the dotted line in the central panel, splits the distribution into the low- and high-\feh\ populations. The \ofe\ histograms of these two populations are shown at the right with the full (black), low-\feh\ (red) and high-\feh\ (dotted blue).
\label{f:mw_ofe}}
\end{figure}

Despite the absence of a bar in the clumpy model, we can compare the vertical distribution of the MDF with the MW's.
In order to do this, we use the model at $6\Gyr$ assuming that the bulge is quenched by bar formation at this time. We apply a coordinate transformation of the model's Cartesian coordinates to Galactic coordinates, $(l, b, d)$, after placing the Galactic center at $8 \kpc$ from the Sun. We  select particles across constant longitude stripes ($-6.5\degrees < l < 6.5\degrees$) at different latitudes ($1.5\degrees < |b| < 2.5\degrees$ and $5.5\degrees < |b| < 6.5\degrees$, restricted to a distance $7<d/\kpc<9$. This represents a typical spectroscopic selection of giant stars in the MW bulge \citep[e.g.][]{wylie+21} with which variations as a function of latitude are studied. Fig.~\ref{f:mw_ofe} shows the resulting distribution of the selected stars in chemical space; these display two over-densities that change their relative contributions as a function of Galactic latitude, as in the MW. 

The chemical track in Fig.~\ref{f:mw_ofe} is comprised of a sequence of \feh-poor stars, followed by a trough and then a shorter sequence of \feh-rich stars whose relative contribution decreases with increasing height from the plane. Without any scaling applied to the simulation, the \feh-rich population is no longer present at a latitude of $b=6\degrees$. Since the simulated galaxy has not formed a bar, the detailed properties of the two populations and their spatial variations are not directly comparable to those in the MW. The specific distributions seen in the MW would depend on many details, such as the epoch of bar formation, the vertical thickening of the bar, and the star and clump formation histories. However, the presence of this overall bimodality and trend in the simulation is consistent with the observations of \citet{rojas-arriagada+19}, \citet{wylie+21} based on [Mg/Fe] abundances from APOGEE and ARGOS data. They showed, from a large number of stars, that the \feh\ bimodality is produced by a low-$\alpha$ sequence of stars over a range of $\sim 0.5$ dex around a solar metallicity that merges with the main high-$\alpha$ sequence. In the simulation, there is a third, much smaller component in Fig.~\ref{f:mw_ofe} that appears as a lower-\ofe\ overdensity in the metal-poor regime. This population becomes more important at higher distances from the plane but clearly remains a minor component throughout. This component is comprised of stars formed in clumps that have lower \ofe\ that form at large radii (see Fig.~\ref{f:chemmapRform}). Their appearance suggests that the clumpy simulation underestimates the evaporation rate of the clumps, possibly because the feedback should be higher. We note, however, that a hint of such a low-\ofe, metal-poor population can be seen in \citet{wylie+21} where an increasing width of the low-\feh\ sequence as a function of height is evident in their figure~25, but further studies with higher number statistics are needed to confirm this.

Despite not having formed a bar, the simulation trends suggest that the chemical bimodality in the bulge produced by clumps is plausibly able to account for the spatial variations of the chemistry of the MW's bulge. 

%%%%%%%%%%%%%%%%%%%%%%%%%%%%%%%%%%%%%%%%%%%%%%%%%%%%%%%%%%%%%%%%%%%%%%%%%%%%%

\section{Discussion} 
\label{s:discussion}

The single track in the chemical space of the bulge, \ie, the absence of an \alfe\ bimodality for a fixed \feh, in the clumpy simulation is similar to that observed in the MW, including the fact that it has two density peaks along the track. Together with the clumpy model's two tracks in the chemical space of the disk (Paper I), this is a striking agreement with the trends seen in the MW, and suggests that the model is capturing a generic behavior. Altogether, these results demonstrate that a holistic view of the chemistry of the entire MW (both bulge and thin$+$thick disks) provides a more stringent constraint on how the early MW formed \citep[see also][]{pdimatteo16}.

\subsection{Comparison with other scenarios}
\label{ss:otherscenarios}

We have shown that an episode of star formation in clumps is able to explain the twin peaks in the bulge's single track in the chemical space. The bulge track follows that of the thick disk at low metallicity but then continues to the thin disk and beyond at high metallicity, as observed in the MW. In Paper I and \citet{beraldoesilva+20} we showed that the chemical thick and thin disks produced via clumps have similar properties (scale-lengths and scale-heights, kinematics, and MDFs) as found in the MW. \citet{amarante+20} showed that clumps also produce the relatively metal-rich population that bridges the thick disk and the inner halo, which has been termed `the Splash' \citep{pdimatteo+19, splash}\footnote{\citet{pdimatteo+19} refer to this feature as ``The Plume''.}. The clump scenario predicts that the thin and thick disks were forming at the same time (Paper I), which appears to be consistent with the presence of RR~Lyrae in the thin disk as well as an age overlap between the chemical thin and thick disks \citep{beraldoesilva+21}. 

Since \gaia's confirmation of the \ges\ (hereafter GSE) \citep{gaia-sausage, enceladus} merger remnant, the chemodynamics of the early MW have been interpreted as products solely of this merger. 
Numerous cosmological simulations have indeed shown that disk chemical bimodalities can arise from gas-rich mergers \citep[e.g.][]{brook+05, snaith+16, grand+18, mackereth+18, buck20}. \citet{mackereth+18} found that such outcomes only occur in about $5\%$ of the EAGLE simulation galaxies, but \citet{buck20} found them to be more common in the NIHAO simulation suite.
Likewise, the Splash has been interpreted as a combination of accreted material and the kinematically heated disk after the GSE merger \citep{pdimatteo+19, splash, mackereth+19, gallart+19}.

The bulge is an important test of the hypothesis that the GSE merger is exclusively responsible for the chemodynamics of the MW because, on the one hand, the merger cannot leave a classical bulge more massive than $\sim 8\%$ of the total stellar mass \citep{jshen+10, bland-hawthorn_gerhard16, debattista+17}, while at the same time producing a bulge chemistry with a single track with two peaks. To date, cosmological simulations that produce the chemical thin and thick disks appear to produce two, or more, tracks in the chemistry of the bulge \citep[e.g.][]{grand+18, buck20}. We have shown here that the chemistry of the bulge can largely, and very naturally, be produced by clumpy star formation (including within the bulge itself). Thus most of the chemodynamics of the early MW, excluding the accreted halo, can now be explained by clumps. However, because the GSE merger certainly happened, it is important to understand to what extent a merger in the presence of clumps is able to explain the details of the MW's chemodynamics. We will be exploring exactly this with project GASTRO \citep{amarante+22}.

Further complicating matters, besides the GSE, there have been suggestions of at least one other equally massive merger in the MW during its early evolution \citep{massari+19, forbes20, horta+21}. \citet{horta+21} used APOGEE DR16 and Gaia DR2 to characterize the stars of this merger event, which they called ``Heracles"\footnote{\citet{massari+19} and \citet{forbes20} dubbed this remnant ``Kraken'' and ``Koala'', respectively.}. They estimated its stellar mass as $\sim 5\times10^8$ $\rm M_{\odot}$, \ie\ as massive as GSE \citep[see also][]{kruijssen+20}. The stars associated with Heracles are located at $R<5~\kpc$, and are thus more bound to the Galactic potential than the GSE remnant \citep[but, see also][for a discussion of whether Heracles could be an artifact in the $E-L_z$ plane of APOGEE's selection function]{lane+22}. These stars are also chemically distinct from the GSE \citep{horta+21, naidu+22} and would imprint as bursts in the SFH of the inner MW \citep{orkney+22}. \citet{naidu+22} estimated it was accreted $\sim 1.7 \Gyr$ before GSE. Recently, \citet{myeong+22} argued for an in-situ origin of Heracles. More recently, this population has been interpreted as the first stars that formed in the MW, based on {\it Gaia}, APOGEE DR17, and H3 survey data \citep{belokurov_kravtsov22, conroy+22, rix+22}.

An alternative, popular model for the formation of the geometric thick
disk posits that it formed in situ, already thick, in an
``upside-down, inside-out'' manner \citep{bird+13, bird+21}. Support
for this model includes the short scale-length of the (chemical) thick
disk \citep{bovy+12b, hayden+15} and the homogeneity of the
high-\al\ population. This scenario is also supported by the high
  gas velocity dispersions and star formation rates in high-redshift
  galaxies \citep[e.g.][]{kassin+12, wisnioski+15}.  The lack of flaring in
the high-\al\ populations is also consistent with the upside-down
scenario (\citet{bovy+16, mackereth+17} but see also
\citet{lian+22}). In general, however, studies of the
upside-down formation scenario have provided no explanation for
the disk chemical bimodality, or the chemistry of the bulge.  An
alternative flavor of the upside-down formation scenario is based on
misaligned star formation. \citet{meng_gnedin21} showed that, in their
cosmological simulations, stars always form in a thin disk, even at
$z>1.5$, and only give the appearance of an upside-down formation
because disks tilt rapidly at early times, which leaves the
star-forming plane misaligned (warped) with respect to the main disk
plane. The subsequent precession of the stars formed off the plane
continuously inflates the height of the main disk \citep[see
  also][]{khachaturyants+21}. More recently, \citet{tamfal+22}
  used a high-resolution ($\sim 10^{9}$ particles) zoom-in
  cosmological simulation to show that the disk is already forming
  thin as early as $z\sim 7-8$, with no upside-down formation. This
  rotationally supported disk thickens slowly due to internal
  instabilities and external perturbations, with stellar accretion
  from satellites providing the main geometric thick disk.
In a similar vein, \citet{vintergatanI} \citep[see also][]{vintergatanII, vintergatanIII} proposed that the origin of the chemical bimodality of the thin$+$thick disks is due to different chemistry in the inner and outer disks which accreted their gas from separate filaments. Early rapid star formation and mergers in the inner disk gave rise to the high-\al\ thick disk population, while star formation in the outer misaligned disk is inhibited by the low density of the gas until the last major merger triggers star formation in the outer disk, which becomes the metal-poor, low-\al\ thin disk. The continuing star formation then builds the metal-rich, low-\al\ thin disk we see today.
\citet{vintergatanII} presented the chemistry of this simulation; the bimodal tracks extend to the inner galaxy, contrary to what is observed in the MW. It is unclear whether this outcome can be avoided in this scenario. 

The classical two-infall scenario of \citet{chiappini+97} \citep[see also][]{chiappini09, bekki_tsujimoto11, tsujimoto_bekki12, grisoni+17, spitoni+21} proposes that the formation of two sequences in the disk chemistry results from two-infall episodes, with high SFR during the first infall, producing the high-\al\ sequence, followed by a second infall with low SFR, producing the low-\al\ sequence. As noted in Paper I the clump model is similar, in terms of SFR, to this model, and the outcome may be indistinguishable. However, our results for the nonclumpy model, which has an SFH not much different from that of the clumpy model, but which fails to form a disk chemical bimodality, is at odds with a pure early high SFR producing a disk chemical bimodality. Clumps produce the chemical bimodality by boosting the star formation rate {\it density} by a factor of $\sim 100$ compared to distributed star formation \citep{clarke+19}. \citet{khoperskov+21} presented several simulations which produced a thin$+$thick disk chemical bimodality which they attributed to the rapidly dropping SFR, coupled with outflows \citep[see also][]{vincenzo_kobayashi20} similar to the two-infall model. The authors also noted that their models undergo a period of clump formation, with comparable clump masses to what we found in Paper I.

\subsection{Observational tests}

Clumps are observed in more than half of high-redshift MW progenitors \citep[e.g.][]{elmegreen_elmegreen05, ravindranath+06, elmegreen+07, forster-schreiber+11, genzel+11, guo+12, guo+15}.  Observed at high resolution, clumps are found to have sizes of order $100-500\pc$, average masses of $\sim 10^8\Msun$ \citep{livermore+12, livermore+15, fisher+17, cava+18} and contribute about $7\%$ of the ongoing star formation rate \citep{wuyts+12}.
Aside from the formation of a geometric thick disk \citep{bournaud+09, clarke+19, beraldoesilva+20}, the presence of clumps does not lead to substantial differences in the morphological properties of galaxies. Indeed the cosmological zoom-in simulations of \citet{inoue_yoshida19}, with identical initial conditions but varying gas physics, found a strong dependence of clump formation on the equation of state of the gas, but very little effect on the global properties of the galaxies.  The signatures of clumps are therefore primarily chemical, because the masses of the clumps are modest \citep{livermore+12, fisher+17, cava+18, benincasa+19}, and the clump epoch lasts only a brief time, until the gas mass fraction declines \citep{cacciato+12}.  We showed in Paper I that the clumps that form in the clumpy simulation are comparable to the ones found in high-redshift galaxies and predicted that chemical bimodalities in disks should be common. Using MUSE spectroscopy, \citet{scott+21} showed that the MW analog UGC~10738 has an \al-rich geometric thick disk, from which they conclude that accretion events are unlikely sources of thick disks. More studies such as this can help establish whether geometric thick disks are \al-enhanced. This will be particularly important for exploring the merger hypothesis, since the merger histories of galaxies are very variable \citep[e.g.][]{lacey_cole93, stewart+08, boylan-kolchin+10}.

Upcoming data from the James Webb Space Telescope will measure the chemistry of the Andromeda galaxy's disk from resolved stellar spectroscopy. Andromeda is known to have had a much more active merger history than that of the MW \citep[e.g.][]{mcconnachie+10, weisz+14, mcconnachie+18, dsouza_bell18, hammer+18}. If clumps played an important role in its chemical evolution, we expect the chemistry of Andromeda's old disks to be comprised of, at least, a high-\al\ and a low-\al\ track somewhat similar to the MW's, with possibly additional merger-induced tracks.

However more detailed tests must necessarily come from the MW since we can study it in much greater detail than any other galaxy. Understanding to what extent the outcome of the GSE merger is degenerate with the clump scenario is an important ingredient in unravelling the formation of the MW. A holistic approach, considering the properties of the bulge, the thin$+$thick disks, and the Splash, is vital to this enterprise. However efforts thus far have been hampered by the relatively small datasets comprising thousands of stars. Future space-based (e.g., \gaia) and ground-based  observatories (e.g. Vera Rubin Telescope) surveys will permit proper-motion measurements of large samples of bulge stars to help unravel the formation of the bulge \citep{gough-kelly+22}. 

A possible test is the distribution of ages at the bulge's low-\feh\ peak versus that of the trough between the two peaks. Most stars in the MW's bulge will now be old; measuring an age difference of $\sim 2\Gyr$ in a present-day $\sim 10\Gyr$-old bulge \citep{ortolani+95, kuijken_rich02, zoccali+03, ferreras+03, sahu+06, clarkson+08, clarkson+11, brown+10, valenti+13, calamida+14, renzini+18, surot+19} is challenging.
Nonetheless, the chemical thin and thick disks do appear to overlap in age, as seen by the existence of RR~Lyrae with small vertical excursions and low \alfe\ \citep{prudil+20}. An age overlap between the MW's thin and thick disk, which was predicted in Paper I has also been demonstrated by \citet{beraldoesilva+21} using the stellar ages of turnoff and giant stars from the \citet{sanders_das18} catalog. Likewise, \citet{silva-aguirre+18} find an age overlap between high-\al\ and low-\al\ disk stars from astroseismic ages. \citet{gent+22} reach a similar conclusion based on data from the \gaia-ESO survey together with \gaia\ EDR3 data. Thus it may well be possible to measure the mean age difference between stars at the trough and those in the low-\feh\ peak to test whether clumps have contributed to the bulge.

\subsection{Clumps as probes of feedback implementations}

Clumps were first proposed to play a role in the formation of
  bulges by \citet{noguchi99}.  Following this suggestion, several
  works explored the role of clumps in bulge formation
  \citep{immeli+04, bournaud+07, elmegreen+08, aumer+10,
    inoue_saitoh12}.  When the cosmological setting is also included,
  the possibility of ``ex-situ'' clumps forming directly in the cold
  gas streaming in before reaching the disk was also recognized
  \citep{dekel+09, ceverino+10}.  Clumps can even be excited by
  external perturbations \citep{inoue+16}.  The cosmological
  simulations of \citet{dubois+21} find that $\sim 10\%$ of the
  stellar mass of $z=4$ galaxies may be in the form of clumps, while
  those of \citet{mandelker+14} resulted in $60\%$ of galaxies hosting
  an in-situ clump population. Meanwhile \citet{mandelker+17} showed
  that bulges can host their own clump, which is more robust to
  feedback; they further showed that radiation pressure increases the
  cold gas fraction (by delaying star formation), increasing the
  lifetime of low-mass clumps.  \citet{inoue_saitoh12} showed that
  bulges formed from clump mergers are rapidly rotating, exponential,
  and comprised of old, metal-rich stars, similar to the bulge of the
  MW. In addition, clumps may further affect the formation of the
  bulge by funnelling gas to the center, leading to further star
  formation and compaction \citep{dekel_burkert14}.

However other studies have questioned the importance of clumps for the evolution of galaxies.  Efficient coupling of feedback energy to gas destroys clumps \citep{elmegreen+08, hopkins+12} and many simulations that employ high feedback prescriptions have failed to find significant clumps or have found ones that do not contribute much to bulges \citep[e.g.][]{tamburello+15}. The short-lived clumps in the FIRE simulations do not manage to migrate to the bulge \citep{oklopcic+17}, and may not even have been bound.  Similarly, in the NIHAO simulation suite, \citet{buck+17} found that clumps are only present in the light, not in the mass, and therefore have minimal contribution to bulge growth. 

The detailed treatment of various forms of feedback
\citep[e.g.][]{fensch_bournaud21} therefore plays an important role
in the ease with which clumps form in simulations.
Moreover, in simulations of single giant molecular clouds (GMCs), the
energy imparted by photoionization, winds, and supernova feedback can
be channeled along preferred directions, thereby preserving the GMC
for a longer time than would otherwise be expected
\citep{rogers_pittard13, dale17, howard+17}.
Thus tests of what role, if any, clumps have played in the evolution
of galaxies like the MW can inform improvements in subgrid
implementations of feedback on the smallest scales, perhaps by
accounting for feedback channeling. Further study of the impact of
clumps on galaxy formation therefore may have much broader impact on
the study of galaxy formation \citep[e.g.][]{dekel+22}. Recently,
\citet{marasco+22} found that the observed outflows from a sample of
starbursting dwarf galaxies are lower than predicted by cosmological
simulations that employ high feedback. They find mass loading factors
of warm gas outflows more than 2 orders of magnitude lower than
predicted, providing strong support for the need for gentler feedback
prescriptions.

\subsection{Caveats}

The simulation presented in this paper is clearly idealized and lacks some of the ingredients that have been suggested to have mattered in the MW's chemical evolution. Of these the most important is the merger of the \ges\ progenitor. The effect of the GSE merger will be explored in future papers \citep[\eg,][]{amarante+22}.

Our simulations place the initial gas in a hot corona. This is
  appropriate for a galaxy of the MW's mass since redshift $z\sim 1$
  \citep{birnboim_dekel03, keres+05}, but is less appropriate before
  then. More realistically, the gas should flow in along filaments
  (cold-mode accretion). Unfortunately setting up such initial
  conditions for high-resolution simulations is difficult.  However
  cosmological simulations have found clumps in galaxies still in the
  filamentary cold accretion mode \citep{dekel+09, ceverino+10};
  provided that the inflow rate of the gas, and the resulting clump
  and star formation rates are realistic, it does not matter how gas
  reaches the disk. If gas stalls in the outer disk \citep[for
    instance as some of the gas does in][]{vintergatanI}, then it may
  be that gas surface densities are never high enough for clumps to
  form. We speculate that if this inhibits the flow of gas to the
  bulge then the bulge itself may never reach the same
  high-\alfe\ state as the thick disk. However in general filamentary,
  cold-mode accretion need not alter the general picture much so long
  as realistic clumps form.

One limitation of the clumpy model presented here is that the clump population in this particular simulation may be too large. This is suggested by high rotation velocity at the center \citep{clarke+19}, and the failure to form a bar (although bars often fail to form at this mass resolution). These effects may be improved in models with higher feedback that still permit long-lived, but lower-mass clumps, which are still able to produce a high-\al\ population (e.g. Garver et al. submitted).

Despite the absence of a bar, we may anticipate what the influence of a bar might be. \citet{debattista+17} showed that many of the trends with  metallicity observed in the MW's bulge can be explained by the secular evolution of the bar, via a mechanism they termed {\it kinematic fractionation}.  In this mechanism, different populations are separated by the bar formation on the basis of their radial random motion.  Populations with large radial random motions are lifted by the bar to large heights ending as a spheroidal population and forming a weaker bar, whereas populations that start with lower radial random motion do not rise to as large heights but end with a more strongly peanut-shaped distribution and a stronger bar. 
As a result, in general the X-shape is better traced by the metal-rich stars, which are younger and start out cooler, while metal-poor stars, which are older and thus kinematically hotter, trace a more boxy structure \citep{debattista+17, athanassoula+17, buck+18, debattista+19, fragkoudi+20}.
Subsequently, \citet{debattista+20} demonstrated that the vertical thickening of stellar populations increases monotonically with the radial action of stars from {\it before} the bar formed. Since stars with larger radial random motion are typically older, and usually more metal-poor, kinematic fractionation results in a vertical metallicity gradient.  In addition the X-shape ends up better traced by metal-rich stars, as is observed in the MW \citep{ness+12, uttenthaler+12, rojas-arriagada+14}. We have shown here that the radial action, $J_R$, decreases along the bulge's chemical track with increasing metallicity. Thus kinematic fractionation would raise stars at the metal-poor peak to larger heights than those of the metal-rich peak. This would further enhance the trends of Section~\ref{ss:mwcomp}, which already match those observed in the MW.

Recently \citet{queiroz+21} derived distances of a large sample of bulge stars with {\sc starhorse} using APOGEE DR16 and \gaia\ EDR3 parallaxes. They argued that the chemistry of the bulge is comprised of not one but two tracks, contrary to earlier studies. The two tracks do not overlap in metallicity (unlike the thin$+$thick disks), but are separated by a gap and have different slopes. After accounting for the stellar population-dependent selection function of APOGEE, \citet{eilers+22}, also found tracks with different slope, although the tracks still do not overlap in metallicity. If these trends are confirmed by imminent large surveys with instruments such as MOONS \citep{moons} and 4MOST \citep{4most}, this may suggest that the clump scenario needs alteration or is perhaps wrong.

\subsection{Summary}

The principal results of this paper are as follows:
\begin{enumerate}
\item A single track with two peaks in the bulge's \feh-\alfe\ space results when clump formation can occur in the early evolution. Clumps sink to the center, contributing to the bulge. The bulge is later populated by more metal-rich, \al-poor stars that form in situ after the epoch of clump formation.  Such twin peaks are not present when the feedback suppresses clump formation.  The relative mass in the high- and low-\feh\ peaks constrains the epoch when star formation in the bulge is quenched (see Sections~\ref{ss:bulgechem}, \ref{ss:chemevol}, and \ref{ss:formlocation}).
\item Star formation within the bulge occurs in the high-\sfrd\ clump mode. This ensures that a separate low-\alfe\ track never forms (see Section~\ref{ss:sfrd}).
\item The metal-rich bulge population, while on average younger than the metal-poor population, overlaps with it in age because the latter population is partly built from stars that came in with clumps after the chemical evolution of in-situ star formation in the bulge had moved to higher metallicities (see Section~\ref{ss:ages}).
\item By the end of the clump epoch, the bulge is already rapidly rotating. The high-\alfe, low-\feh\ bulge population is kinematically hotter than the low-\alfe, high-\feh\ one (see Section~\ref{ss:alphakine}).
\item The population at the metal-rich peak is prominent at low latitudes but declines with distance from the mid-plane, as observed in the MW (see Section~\ref{ss:mwcomp}).
\item A test of the role of clumps on the MW's bulge comes from a comparison of the age distributions of the low-\feh\ peak and the trough between the two peaks. In the presence of clumps, the age distributions overlap significantly, with the mean age higher in the trough than at the low-\feh\ peak, contrary to the usual expectation of increasing metallicity with age (see Section~\ref{ss:ages}). 
\end{enumerate}

This paper, together with Paper I, presents idealized simulations that demonstrate that clump formation provides a very direct and natural way of producing chemical trends observed not only in the MW's thin$+$thick disk, but also in the bulge. The simulations are by no means wholly realistic, but the ease with which they produce the trends observed in the MW encourages us to explore further the role of clumps in the early history of galaxies. In contrast, satisfying both constraints in other scenarios of thick disk formation may require a more specific set of circumstances, which would mean the MW is unusual.  The clump model makes some important predictions that can be verified with future facilities, including that chemical thick disks should be common in MW-mass galaxies and that a population of chemical thin-disk stars of comparable age to the thick disk should exist in the MW. Studying the consequences of the clumps in such simulations may also provide a useful probe of feedback implementations.

%%%%%%%%%%%%%%%%%%%%%%%%%%%%%%%%%%%%%%%%%%%%%%%%%%%%%%%%%%%%%%%%%%%%%%%%%%%%%

\bigskip
\noindent
{\bf Acknowledgements.}

\noindent
V.P.D., L.B.S., and T.K. were supported by STFC Consolidated grant ST/R000786/1.  D.J.L. was supported for part of this project by a UCLan UURIP internship. L.B.S acknowledges the support of NASA-ATP award 80NSSC20K0509 and Science Foundation AAG grant AST-2009122. J.A.S.A. acknowledges funding from the European Research Council (ERC) under the European Union’s Horizon 2020 research and innovation program (grant agreement No. 852839). M.Z. acknowledges support from the ANID BASAL Center for Astrophysics and Associated Technologies (CATA) through grants AFB170002, ACE210002, and FB210003, the ANID Millennium Institute of Astrophysics (MAS) ICN12\_009 and ANID Fondecyt Regular grant 1191505. E.V. acknowledges the Excellence Cluster ORIGINS funded by the Deutsche Forschungsgemeinschaft (DFG; German Research Foundation) under Germany’s Excellence Strategy \-- EXC\--2094\--390783311. S.A.K. would like to acknowledge support from NASA’s Astrophysics Data Analysis Program (ADAP) grant number 80NSSC20K0760.
We thank the anonymous referee for comments that helped improve this paper.
An important part of the methodology for the stellar population modeling used in this paper was worked out in 2018 at the Aspen Center for Physics, which is supported by National Science Foundation grant PHY-1607611. The visit of V.P.D. was partially supported by a grant from the Simons Foundation.  The simulations in this paper were run at the DiRAC Shared Memory Processing system at the University of Cambridge, operated by the COSMOS Project at the Department of Applied Mathematics and Theoretical Physics on behalf of the STFC DiRAC HPC Facility: www.dirac.ac.uk. This equipment was funded by BIS National E-infrastructure capital grant ST/J005673/1, STFC capital grant ST/H008586/1 and STFC DiRAC Operations grant ST/K00333X/1. DiRAC is part of the National E-Infrastructure.  This paper is dedicated to the memory of George Lake, for whose support and inspiration V.P.D. is deeply indebted.


% \bibliography{allrefs}
% \bibliographystyle{aasjournal}
%\bibliography{allrefs}
\bibliography{main}
\bibliographystyle{aasjournal}


\end{document}